\documentclass[a4paper,11pt]{article}
\pdfoutput=1 

\usepackage{jheppub}
\usepackage[T1]{fontenc} 

\usepackage{tensor}
\usepackage{amsmath,amsfonts,amssymb}
\usepackage[mathscr]{euscript} 
\usepackage{graphicx, enumerate}
\usepackage[dvipsnames]{xcolor}

\usepackage[normalem]{ulem}

\DeclareFontEncoding{LS1}{}{}
\DeclareFontSubstitution{LS1}{stix}{m}{n}
\DeclareSymbolFont{stixsymbols}{LS1}{stixscr}{m}{n}
\SetSymbolFont{stixsymbols}{bold}{LS1}{stixscr}{b}{n}
\DeclareMathSymbol{\kay}{\mathalpha}{stixsymbols}{"6B}
\DeclareMathSymbol{\hay}{\mathalpha}{stixsymbols}{"68}
\newcommand\CC[1]{\mathfrak{#1}}

\newcommand{\overbar}[1]{\mkern 1.5mu\overline{\mkern-1.5mu#1\mkern-1.5mu}\mkern 1.5mu}

\newcommand{\TT}{T\overbar{T}}
\newcommand{\JJ}{J\overbar{J}}
\newcommand{\de}{\text{d}}

\newcommand{\alg}[1]{\mathfrak{#1}}
\newcommand{\PCM}{\text{PCM}}
\newcommand{\PCMWZ}{\text{PCMWZ}}
\newcommand{\SSSM}{\text{SSSM}}
\newcommand{\sSSSM}{\text{sSSSM}}
\newcommand{\jr}{\tilde{\jmath}}

\DeclareMathOperator{\ch}{\text{ch}}
\DeclareMathOperator{\sh}{\text{sh}}

\DeclareMathOperator{\tr}{\text{tr}}
\DeclareMathOperator{\str}{\text{str}}
\DeclareMathOperator{\Ad}{\text{Ad}}

\title{\boldmath On the Classical Integrability of Root-$\TT$ Flows}

\author[a]{Riccardo Borsato,}
\author[b]{Christian Ferko,}
\author[c,d,1]{Alessandro Sfondrini\note{IBM Einstein Fellow.}}

\affiliation[a]{Instituto Galego de F\'isica de Altas Enerx\'ias (IGFAE) and Departamento de F\'isica de Part\'iculas,\\
Universidade de Santiago de Compostela}
\affiliation[b]{Center for Quantum Mathematics and Physics (QMAP), 
\\ Department of Physics \& Astronomy,  University of California, Davis, CA 95616, USA}
\affiliation[c]{Dipartimento di Fisica e Astronomia, \\
Universita degli Studi di Padova, \\
\& Istituto Nazionale di Fisica Nucleare, \\
Sezione di Padova, via Marzolo 8, 35131 Padova, Italy.}
\affiliation[d]{Institute for Advanced Study \\
Einstein Drive, Princeton, New Jersey, 08540 USA.}

\emailAdd{riccardo.borsato@usc.es}
\emailAdd{caferko@ucdavis.edu}
\emailAdd{alessandro.sfondrini@unipd.it}

\abstract{%
The Root-$\TT$ flow was recently introduced as a universal and classically marginal deformation of any two-dimensional translation-invariant field theory. The flow commutes with the (irrelevant) $\TT$ flow and it can be integrated explicitly for a large class of actions, leading to non-analytic Lagrangians reminiscent of the four-dimensional Modified-Maxwell theory (ModMax).
It is not a priori obvious whether the Root-$\TT$ flow preserves integrability, like it is the case for the $\TT$ flow.
In this paper we demonstrate that this is the case for a large class of classical models by explicitly constructing a deformed Lax connection. We discuss the principal chiral model and the non-linear sigma models on symmetric and semi-symmetric spaces, without or with Wess-Zumino term. We also construct Lax connections for the two-parameter families of theories deformed by both Root-$\TT$ and $\TT$ for all of these models.
}

\begin{document}
\maketitle
\flushbottom


\newpage
\section{Introduction}
\label{sec:introduction}

Given the action $\mathcal{S}_0$ of a quantum field theory (QFT) and an integrated local operator $\mathcal{O}$, it is possible to formally deform the action infinitesimally by setting
\begin{equation}
\label{eq:generic-flow}
    \mathcal{S}_{\lambda} = \mathcal{S}_{0}+ \lambda\, \mathcal{O} + O(\lambda^2)\,. 
\end{equation}
For instance, in the case where $\mathcal{S}_{0}$ describes a conformal field theory (CFT) and $\mathcal{O}$ is a \textit{relevant} operator in the spectrum of the theory, it is natural to interpret $\mathcal{S}_{0}$ as an ultraviolet (UV) theory, and $\mathcal{O}$ as the source of a renormalisation group (RG) flow. Geometrically, $\mathcal{O}$ is a tangent vector to the RG flow at a fixed point, and the RG flow equations determine the remainder of the $\lambda$-expansion in~\eqref{eq:generic-flow}. Determining the flow completely is a very difficult problem in general. Remarkably, in two-dimensions it is sometimes possible to investigate such flows quite explicitly in the special setup where the perturbing operator preserves ``sufficiently many symmetries''. In this case, the flow is integrable and can be studied \textit{e.g.}\ by thermodynamic Bethe ansatz techniques, see \textit{e.g.}~\cite{Zamolodchikov:1989cf,Zamolodchikov:1991vx,Mussardo:1992uc,Bazhanov:1994ft}.

For two-dimensional QFTs it is possible to construct other deformations quite explicitly. If $\mathcal{S}_0$ enjoys at least two conservation laws with currents $j^\mu_{0}$ and $k^\mu_0$,
\begin{equation}
    \partial_\mu j^\mu_0 = \partial_\mu k^\mu_0=0\,,
\end{equation}
then we may define the combination
\begin{equation}
    \mathcal{O}_{0}^{j\wedge k} = \int \de\tau\int \de\sigma\;  j^\mu_0(\tau,\sigma)\, k^\nu_0(\tau,\sigma)\,\epsilon_{\mu\nu}\,.
\end{equation}
It turns out that this composite operator can be defined in the quantum theory without incurring any short-distance singularities~\cite{Smirnov:2016lqw}. Then, it is possible to define a deformed action by the equation
\begin{equation}
\label{eq:current-current}
    \partial_{\lambda} \mathcal{S}_{\lambda} = \mathcal{O}_{\lambda}^{j\wedge k}\,,
\end{equation}
where $\mathcal{O}_{\lambda}^{j\wedge k}$ is defined in the deformed theory, out of the deformed currents $j^\mu_\lambda$ and  $k^\mu_\lambda$ (which are conserved with respect to $\mathcal{S}_{\lambda}$). This is a rather general definition that encompasses many well-studied examples, including the \textit{marginal} $\JJ$ deformation (when $J^\mu=j^\mu$ and $\epsilon^{\mu\nu}J_\nu=k^\mu$ are both conserved), and the \textit{irrelevant} $\TT$ deformation~\cite{Zamolodchikov:2004ce,Smirnov:2016lqw,Cavaglia:2016oda}, which arises when $T^{0\mu}=j^\mu$ and $T^{1\mu}=k^\mu$. 
In that case
\begin{equation}
    \mathcal{O}_{\lambda}^{(2)} = - \int \de\tau\int \de\sigma\;  \det\left[T^{\mu\nu}_{\lambda}(\tau,\sigma)\right]\,.
\end{equation}
The $\TT$ deformation is universal: it can be constructed for any translation invariant QFT. Remarkably, it can be studied very explicitly both at the classical and at the quantum level.
Moreover, if the seed theory $\mathcal{S}_0$ happens to be a CFT, or an integrable QFT (IQFT), infinitely-many analogues of $\TT$ may be constructed~\cite{Smirnov:2016lqw}: both CFTs and IQFTs possess infinitely-many higher-spin conserved currents, which may be used to define \textit{more and more irrelevant}  Poincar\'e-invariant operators of the form~\eqref{eq:current-current}. Namely, the currents related to spin $\pm s$ operators define an operator~$\mathcal{O}^{s}$ of dimension $(2s-2)$.
Very remarkably, each of these operators~$\mathcal{O}^{(s)}$, including $\TT$, defines a flow that preserves all higher-spin charges~\cite{Smirnov:2016lqw,Hernandez-Chifflet:2019sua}. In other terms, $\TT$ and its higher-spin counterparts \textit{preserve integrability}. This allows for a quite detailed study of these deformations, \textit{e.g.}\ yielding flow equations for various physical quantities along the deformation, most notably the finite-volume energy spectrum~\cite{Smirnov:2016lqw,Cavaglia:2016oda}.

Recently, a new type of \textit{classically marginal} deformation of two-dimensional field theories has been considered~\cite{Ferko:2022lol}, see also~\cite{Conti:2022egv,Babaei-Aghbolagh:2022kfz}. Here, the deforming operator is taken to be%
\footnote{In what follows, we will always denote the parameter of the Root-$\TT$ flow as $\gamma$, in contrast with the parameter~$\lambda$ of a generic deformation. The parameter of the irrelevant $\TT$ deformation will also be indicated by~$\lambda$.}
\begin{equation}
    \mathcal{R}_{\gamma} = - \int \de\tau\int \de\sigma\;  \sqrt{\det\left[\widetilde{T}^{\mu\nu}_{\gamma}(\tau,\sigma)\right]}\,,\qquad
    \widetilde{T}^{\mu\nu}_{\gamma}=T^{\mu\nu}_{\gamma}-\frac{1}{2}g^{\mu\nu}\,T^{\rho\sigma}_{\gamma}g_{\rho\sigma}\,,
\end{equation}
where $g_{\mu\nu}$ is the two-dimensional metric.
In light-cone coordinates%
\footnote{%
We take $g_{\pm\mp}=-2$ and $g^{\pm\mp}=-\tfrac{1}{2}$ so that for any vector $V^\mu$, $V^\pm = \tfrac{1}{2}(V^0\pm V^1)$ and $V_{\pm}=V_0\pm V_1$.
}
\begin{equation}
    \mathcal{R}_{\gamma} = \int \de\sigma_+\int\de\sigma_-\; \sqrt{\widetilde{T}^{++}_\gamma(\sigma_+,\sigma_-)\,\widetilde{T}^{--}_\gamma(\sigma_+,\sigma_-)}\,.
\end{equation}
It is worth emphasising that $\widetilde{T}^{\pm\pm}$ are chiral only if $\mathcal{S}_{\gamma}$ defines a CFT --- otherwise they genuinely depend on both $\sigma_{+}$ and $\sigma_{-}$.
The classical flow is generated by setting
\begin{equation}
    \partial_\gamma \mathcal{S}_\gamma = \mathcal{R}_\gamma\,,
\end{equation}
and it has has several intriguing properties~\cite{Ferko:2022lol}:
\begin{enumerate}
    \item It preserves classical conformal invariance, \textit{i.e.}\ if the stress-energy tensor of the seed theory obeys $T^{\mu\nu}_0 g_{\mu\nu}=0$, then in the deformed theory $T^{\mu\nu}_\gamma g_{\mu\nu}=0$.
    \item It commutes with the $\TT$ flow generated by $\mathcal{O}^{(2)}$, which allows us to define a two-parameter flow resulting in an action $\mathcal{S}_{(\lambda,\gamma)}$.
    \item If $\mathcal{S}_0$ is given by a member of a large class of Lagrangian densities $L_0$, it is possible to express $L_{\gamma}$ (as well as $L_{(\lambda,\gamma)}$) in closed~form.
    \item The resulting deformed Lagrangian is not analytic in a derivative expansion, and is very reminiscent of that of modified Maxwell theories in four dimensions~\cite{Bandos:2020jsw,Bandos:2020hgy}. 
\end{enumerate}
To illustrate the last two points, let us consider as seed action the one for $N$ non-interacting massless fields $\phi^i$, $i=1,\dots N$,
\begin{equation}
    \mathcal{S}_0 = \int\de\tau\int\de\sigma \; \frac{1}{2}\partial_\mu \phi^i\partial^\mu \phi^i\,.
\end{equation}
Then we have~\cite{Ferko:2022lol,Conti:2022egv}
\begin{equation}
\label{eq:freedeformed}
    \mathcal{S}_\gamma =  \int\de\tau\int\de\sigma \left(\frac{\cosh\gamma}{2}\partial_\mu \phi^i\partial^\mu \phi^i+\frac{\sinh\gamma}{2} \sqrt{2\partial_\mu \phi^i\partial^\nu \phi^i\partial_\nu \phi^j\partial^\mu \phi^j
    -\left(\partial_\mu \phi^i\partial^\mu \phi^i\right)^2}\right).
\end{equation}
With the exception of the case of the single boson, when
\begin{equation}
    \mathcal{S}_\gamma = e^\gamma\, \mathcal{S}_{0}\,,\qquad(N=1),
\end{equation}
the action is non-analytic for small gradients~$\partial_\mu \phi^i$ and in fact quite intricate. For instance, although it still admits plane waves such as $\phi^\mu=n^\mu e^{i p_\nu \sigma^\nu}$ as solutions, \textit{linear combinations} of plane waves are not solutions. This is very similar to ModMax theories~\cite{Bandos:2020jsw,Bandos:2020hgy,Bandos:2021rqy,Lechner:2022qhb} (see~\cite{Sorokin:2021tge} for a review) and in fact the action~\eqref{eq:freedeformed} can be obtained as the dimensional reduction of the ModMax action~\cite{Conti:2022egv}.
This non-linear and non-analytic structure makes the quantisation of Root-$\TT$ deformed theories, as well as of ModMax theories, an outstanding problem. This is in stark contrast with $\TT$ deformed theories where at least some features of the model -- like the spectrum \cite{Smirnov:2016lqw,Cavaglia:2016oda}, S matrix \cite{Dubovsky:2017cnj}, partition function \cite{Cardy:2018sdv,Datta:2018thy,Aharony:2018bad,Griguolo:2022xcj,Griguolo:2022hek,Griguolo:2022vdx}, and correlation functions \cite{Cardy:2019qao,Kraus:2018xrn} -- may be studied quite thoroughly at the quantum level.

In an effort to better understand Root-$\TT$ theories (and perhaps ModMax theories too), it is natural to ask if this deformation has any further property, at least at the classical level. In particular, does it preserve integrability, like it is the case for $\TT$ and its higher-spin analogues generated by $\mathcal{O}^{(s)}$, and how can this be seen?

In this work we will not answer this question in full generality, but instead give evidence for a (rather large) class of theories. These are the principal chiral model (PCM) on a Lie group~$G$, and two classes of non-linear sigma models (NLSMs) on coset spaces $G/H$, where $G$ is a Lie (super)group and $H$ a bosonic subgroup thereof, as well as the extensions of these models by a Wess-Zumino (WZ) term. These NLSMS are the (bosonic) symmetric-space sigma model (SSSM) and its supersymmetric extension, the ``semi-symmetric space'' sigma model (sSSSM). For all these models we construct the deformed Lax connection $\CC{L}_\mu$, and check that imposing  its flatness
\begin{equation}
    \partial_+\CC{L}_{-}-\partial_-\CC{L}_{+} + \left[\CC{L}_{+},\,\CC{L}_{-}\right]=0\,,
\end{equation}
is equivalent to the equations of motion of the deformed model. We also construct the Lax connection for the $(\gamma,\lambda)$ doubly-deformed  models, where we combine $\TT$ and Root-$\TT$ deformations. Our construction follows a rather explicit case-by-case analysis, similarly to the construction of $\TT$-deformed Lax connections of~\cite{Chen:2021aid}.

It should be noted that the semi-symmetric space sigma model reduces to the symmetric space sigma model by setting its odd (fermionic) part to zero. In turn, the PCM can also be thought of as a symmetric space coset $(G\times G)/G$. In a sense, we might as well only describe the most general sSSSM case (with the addition of a WZ term), and all the others would follow from it. However, especially given that the construction of the sSSSM is somewhat technical, we find it clearer to first work out in detail the PCM, which is substantially simpler, and then ``work our way up'' to the SSSM and sSSSM.

Let us stress that the identification of a Lax connection is only the first step in the proof of classical integrability of the deformed models. One should further check that the conserved charges constructed from the monodromy matrix have mutually vanishing Poisson brackets, i.e.~they are in involution. In this regard, the models that we are deforming are known to possess some technical complications due to a non-ultralocal Poisson structure. That means that when taking the Poisson brackets not only the Dirac delta function appears but also its derivative. Nevertheless, it is still possible to recast the Poisson brackets of the Lax connection in a useful form known as Maillet brackets~\cite{Maillet:1985ec,Maillet:1985ek} that allows one to construct the mutually conserved charges and argue for the classical integrability of the models. We expect the deformed models analysed here to display a modification of the original structure of Maillet brackets of the undeformed models. It would be very important, and at the same time very interesting, to construct the Maillet brackets explicitly.

This paper is structured as follows. We begin in Section~\ref{sec:review} by reviewing the constructions of Root-$\TT$ flows presented in~\cite{Ferko:2022lol} to the extent that we will need later in the paper. Next, in Section~\ref{sec:pcmwz} we discuss the case of the PCM (possibly with WZ term). We actually discuss the case of the deformation of the PCM in quite some detail (Section~\ref{sec:pcmwz:deform}) as it illustrates nicely the mechanism by which integrability is preserved in the other models too, but it is less involved technically. Adding the WZ term also allows us to consider explicitly the deformation of the classical WZW model, see Section~\ref{sec:pcmwz:wzw}.
In Section~\ref{sec:nlsm} we discuss first the case of symmetric-space sigma models and later that of semi-symmetric space sigma models. Even though the constructions get more and more involved, the logic follows quite closely the one we illustrate for the PCM in the previous section.
We conclude and present some open questions in Section~\ref{sec:conclusions}.
The derivation of the equations of motion for the various (deformed) models of interest, which is straightforward but a little tedious, is presented in Appendix~\ref{app:eom}.

\section{\boldmath Review of the Root-\texorpdfstring{$\TT$}{TTbar} Flows}
\label{sec:review}

Let us briefly review the construction of the deformed Lagrangian presented in~\cite{Ferko:2022lol}, and fix some notation which we will use later in the paper.
To begin with, we are interested in a Lagrangian featuring~$N$ interacting Bosons $\phi^i$, $i=1,\dots N$, of the form
\begin{equation}
\label{eq:bosonicseed}
    L_0 = \frac{1}{2}\left[g^{\mu\nu}G_{ij}(\phi)+\varepsilon^{\mu\nu}B_{ij}(\phi)\right] \partial_\mu\phi^i\partial_\nu\phi^j - V(\phi)\,,
\end{equation}
where $g^{\mu\nu}$ is the two dimensional inverse-metric, which in light-cone coordinates reads $g^{\pm\mp}=-\tfrac{1}{2}$ (while $g_{\pm\mp}=-2$) and $\varepsilon^{\mu\nu}$ is the Levi-Civita tensor with $\varepsilon^{+-}=-\varepsilon^{-+}=\tfrac{1}{2}$. Notice that the potential term $V(\phi)$ will not be needed for the models which we study below, but it turns out that it is easy to include it.

For the purpose of evaluating the Hilbert\footnote{The Root-$\TT$ deformation may also be defined for non-relativistic theories by using the Noether stress tensor. However, in this work we will restrict attention to theories which enjoy boost invariance.} stress-energy tensor
\begin{equation}\label{hilbert_stress_def}
    T_{\mu\nu}= -2\frac{\partial L}{\partial g^{\mu\nu}}+g_{\mu\nu}L\,,
\end{equation}
it is important to single out the terms in~\eqref{eq:bosonicseed} which couple to the two-dimensional metric~$g_{\mu\nu}$. Hence we define
\begin{equation}
    (X_1)_{\mu}{}^{\nu} = G_{ij}(\phi)\,\partial_\mu\phi^i\partial^\nu\phi^j\,.
\end{equation}
When constructing powers of the stress-energy tensor, we will generate new tensor structures by contracting $(X_1)_{\mu}{}^{\nu}$ with itself and with the metric in various ways. Due to the symmetry of the stress-energy tensor and of the metric, it is immediate to see that these terms will be of the form
\begin{equation}
    (X_2)_{\mu}{}^{\nu} = (X_1)_{\mu}{}^{\rho}(X_1)_{\rho}{}^{\nu}\,,\qquad
    (X_3)_{\mu}{}^{\nu} = (X_2)_{\mu}{}^{\rho}(X_1)_{\rho}{}^{\nu}\,,\qquad\dots\,,
\end{equation}
or generally
\begin{equation}
    (X_{n+1})_{\mu}{}^{\nu} = (X_{n})_{\mu}{}^{\rho}(X_1)_{\rho}{}^{\nu}\,,\qquad n\geq 1\,.
\end{equation}
The deformed action will be a scalar and therefore will depend on the traces of the various $(X_n)_{\mu}{}^{\nu}$, which we denote by
\begin{equation}
    x_n = (X_{n})_{\mu}{}^{\mu}\,,\qquad n\geq1\,.
\end{equation}
The deformed Lagrangian will also be a function of the terms in~\eqref{eq:bosonicseed} which do not couple to the metric, which  we  denote by
\begin{equation}
\label{eq:x0}
    x_0 = \frac{1}{2}\varepsilon^{\mu\nu}B_{ij}(\phi) \partial_\mu\phi^i\partial_\nu\phi^j - V(\phi)\,.
\end{equation}
Finally, we note that many of the trace terms $x_n$, $n\geq1$ are related by trace identities. In fact, $(X_n)_{\mu}{}^{\nu}$ is a $2\times2$ matrix and as such we may express any trace term as a polynomial in $x_1$ and $x_2$,
\begin{equation}
    x_n = P_n(x_1,x_2)\,,\qquad n\geq 3\,.
\end{equation}
This is sufficient to conclude that the deformed Lagrangian takes the form
\begin{equation}
    L_{\gamma}(x_0,x_1,x_2)\,,
\end{equation}
with the seed action being 
\begin{equation}
    L_{0}(x_0,x_1,x_2)\, = \frac{1}{2}x_1+x_0\,.
\end{equation}

Once the action is recast in this form, an explicit computation~\cite{Ferko:2022lol} allows us to integrate the flow equation obtaining simply
\begin{equation}
    L_{\gamma}(x_0,x_1,x_2)
    =L_{}(x_0,x_1^{(\gamma)},x_2^{(\gamma)})\,.
\end{equation}
In other words, the deformed Lagrangian takes the same form as the original Lagrangian in terms of the deformed quantities  $x_1^{(\gamma)}$ and $x_2^{(\gamma)}$ (while $x_0$ remains undeformed). These are given by
\begin{equation}
\label{eq:x1x2generic}
\begin{aligned}
    x_1^{(\gamma)}\,&=\ch(\gamma) x_1+\sh(\gamma)\sqrt{2x_2-x_1^2}\,,\\
    x_2^{(\gamma)}\,&=\ch(2\gamma)x_2+\sh(2\gamma)x_1\sqrt{2x_2-x_1^2}\,.
\end{aligned}
\end{equation}
In conclusion, for the Bosonic models below, the deformed Lagrangian will be simply
\begin{equation}
\label{eq:deformedlagrangian}
    L_\gamma = \frac{1}{2}x_1^{(\gamma)}+x_0=
    \frac{\ch(\gamma)}{2} x_1+\frac{\sh(\gamma)}{2}\sqrt{2x_2-x_1^2}+x_0\,.
\end{equation}
It is also worth noting that the combination appearing under the square roots factorises when expressed in light-cone coordinates,
\begin{equation}
    2x_2-x_1^2 = \left(G_{ij}(\phi) \partial_+\phi^i\partial_+\phi^j\right)\left(G_{kl}(\phi) \partial_-\phi^k\partial_-\phi^l\right),
\end{equation}
which will be useful later on. Finally we remark that there exists a conserved quantity along this flow, namely
\begin{equation}
    (x_1^{(\gamma)})^2- x_2^{(\gamma)} =  (x_1)^2-x_2 \,,
\end{equation}
which can be used to show that this flow preserves tracelessness of the stress-energy tensor~\cite{Ferko:2022lol}.
This discussion can also be extended to models with Bosons and Fermions~\cite{Ferko:2022lol}. However, we will not need the details of the fermionic constructions in what follows. In fact, the only model where we might expect complications due to the presence of Fermions is the semi-symmetric space sigma model. In that case, however, the fermionic currents do not couple to the metric (a Fermionic bilinear appears in $x_0$), while the rest of the dependence on Fermions is packaged inside bosonic currents, as we shall see below. 

\subsection{Combining \texorpdfstring{$\TT$}{TTbar} and Root-\texorpdfstring{$\TT$}{TTbar} Deformations}
\label{sec:review:doubledef}

Since the Root-$\TT$ flow commutes with the (irrelevant) $\TT$ flow, it is easy to define a joint deformation which depends on two parameters: $\gamma$ for Root-$\TT$ deformations, and $\lambda$ for $\TT$ deformations~\cite{Ferko:2022lol}.
In fact, it is possible to construct a doubly-deformed Lagrangian, $L_{(\lambda,\gamma)}$ so that
\begin{equation}\label{eq:doubly_deformed_def}
    \partial_{\lambda} L_{(\lambda,\gamma)} = -\text{det}\left[T^{\mu\nu}_{(\lambda,\gamma)}\right],\qquad
    \partial_{\gamma} L_{(\lambda,\gamma)} = \sqrt{\widetilde{T}^{++}_{(\lambda,\gamma)} \widetilde{T}^{--}_{(\lambda,\gamma)}}\,.
\end{equation}
A simple way to do so is to first perform a $\TT$ deformation, and then deform the resulting action along the lines illustrated above. The actions resulting from $\TT$ flows can be expressed in closed form~\cite{Bonelli:2018kik}. To do so, it is convenient to restrict to a slightly more narrow Lagrangian than~\eqref{eq:bosonicseed}. In fact, while the Root-$\TT$ flow is completely insensitive to~$x_0$ of eq.~\eqref{eq:x0} this is not the case for~$\TT$. In particular, the potential $V(\phi)$ enters the flow, which is described by the Burgers' equation, as an initial condition; this results in a fairly intricate dependence on $V(\phi)$ at finite~$\lambda$. Conversely, the topological term is left unchanged by any continuous deformation. For the class of models that we want to consider here, it suffices to restrict to $V(\phi)=0$, \textit{i.e.}
\begin{equation}
    x_0=\frac{1}{2}\varepsilon^{\mu\nu}B_{ij}(\phi)\partial_\mu\phi^i\partial_\nu\phi^j\,,\qquad
    x_1=g^{\mu\nu}G_{ij}(\phi)\partial_\mu\phi^i\partial_\nu\phi^j\,,
\end{equation}
In this case, starting from a seed Lagrangian
\begin{equation}
    L_0(x_0,x_1,x_2) = \frac{1}{2}x_1+x_0\,,
\end{equation}
we have the $\TT$-deformed Lagrangian~\cite{Bonelli:2018kik}
\begin{equation}
\label{eq:TTbardeformedL}
    L_{(\lambda,0)}(x_0,x_1,x_2)=\frac{1}{2\lambda}\left(
    \sqrt{1+2\lambda x_1 +2\lambda^2\big((x_1)^2-x_2\big)}-1\right)+x_0\,.
\end{equation}
Then, according to our general recipe,
\begin{equation}
     L_{(\lambda,\gamma)}(x_0,x_1,x_2)=
     L_{(\lambda,0)}(x_0,x_1^{(\gamma)},x_2^{(\gamma)}),
\end{equation}
or more explicitly
\begin{equation}
\label{eq:doubleformedL}
     L_{(\lambda,\gamma)}(x_0,x_1,x_2)
     = \frac{1}{2\lambda}\left(
\sqrt{1+2\lambda x_1^{(\gamma)}+2\lambda^2\left((x_1^{(\gamma)})^2-x_2^{(\gamma)}\right)}-1
    \right)+x_0,
\end{equation}
with $x_1^{(\gamma)}$ and $x_2^{(\gamma)}$ as in~\eqref{eq:x1x2generic}.

\section{The Principal Chiral Model with Wess-Zumino Term}
\label{sec:pcmwz}
The Principal Chiral Model (PCM) is a particularly simple sigma model where the target space is a Lie group~$G$.  
In this section we briefly review the construction of the Lax connection for the PCM (see also the recent review~\cite{Hoare:2021dix}), and then we extend it to the Root-$\TT$ deformed model. We later show that similar arguments apply also to the case with a Wess-Zumino (WZ) term and to the double Root-$\TT$ and $\TT$ deformation.

\subsection{The Principal Chiral Model}
\label{sec:pcmwz:pcm}
Consider a Lie group $G$ and its Lie algebra $\alg{g}$.
The left- and right-invariant Maurer-Cartan forms are
\begin{equation}
    j = g^{-1}\de g\,,\qquad \jr =- (\de g)g^{-1}\,,
\end{equation}
with $j\in\alg{g}$ and $\jr\in\alg{g}$.
These are related by the adjoint action $\jr=-\Ad_g{j}=-g^{-1}jg$.
Let $g= g(\tau,\sigma)\in G$ be a $G$-valued field. 
We will drop the $(\tau,\sigma)$ dependence to keep our notation lighter. Define $j_\mu$  and $\jr_\mu$ to be the pull-back of the Maurer-Cartan forms
\begin{equation}
    j_\mu = g^{-1}\partial_\mu g\,,\qquad \jr_\mu =- \left( \partial_\mu g \right) g^{-1}\,,
\end{equation}
which both take values in~$\alg{g}$.
Here $j_\mu$ is invariant under left-multiplication by a \textit{constant} group element $g_0\in G$, $g'=g_0\,g$, and $\jr_\mu$ is invariant under right-multiplication by a constant group element, $g'=g\,g_0$.
Both $j_\mu$ and $\jr_\mu$ satisfy the flatness condition, which we express in light-cone coordinates,
\begin{equation}
\label{eq:maurercartan}
    \partial_+ j_- -\partial_- j_+ + \left[j_+,\,j_-\right]=0=\partial_+ \jr_- -\partial_- \jr_+ + \left[\jr_+,\,\jr_-\right] \,.
\end{equation}
The Lagrangian of the principal chiral model takes a simple form in terms of $j_\mu$, namely
\begin{equation}
\label{eq:Lpcm}
    L_{\PCM} = \frac{1}{2}g^{\mu\nu}\tr\left[j_\mu\,j_\nu\right]=-\frac{1}{2}\tr\left[j_+\,j_-\right]\,,
\end{equation}
which can be equivalently written as
\begin{equation}
    L_{\PCM} = \frac{1}{2}g^{\mu\nu}\tr\left[\jr_\mu\,\jr_\nu\right]=-\frac{1}{2}\tr\left[\jr_+\,\jr_-\right]\,.
\end{equation}
Both expressions are manifestly invariant under left-multiplication  and  right-multiplication by $g_0\in G$, hence the action is invariant under~$G\times G$.

As it is easy to verify explicitly, the Noether current associated to the invariance under right-multiplication is $j_\mu$, and its conservation equation
\begin{equation}
\label{eq:eompcm}
    \partial_\mu j^\mu =0\,,
\end{equation}
is equivalent to the equations of motion of the model, see also Appendix~\ref{app:eomPCM} for the explicit computation. Therefore, we may define the Lax connection
\begin{equation}
\label{eq:Laxpcm}
    \CC{L}_{\pm} = \frac{j_\pm}{1\mp z}\,,
\end{equation}
which depends on the spectral parameter $z\in\mathbb{C}$.
Requiring $\CC{L}_\mu$ to be flat for any~$z$ amounts to imposing the equations of motions of the model~\eqref{eq:eompcm}
\begin{equation}
\begin{aligned}
    0\,&= \partial_{+}\CC{L}_- - \partial_{-}\CC{L}_+ + \left[\CC{L}_+,\CC{L}_-\right]\\
    &=\frac{1}{1-z^2}\Big(
    \partial_{+}j_- - \partial_{-}j_+ + \left[j_+,j_-\right]
    -z\left(\partial_+j_- + \partial_- j_+\right)\Big),
\end{aligned}
\end{equation}
as it may be verified explicitly by using the Maurer-Cartan equation~\eqref{eq:maurercartan}.
Similarly, the Noether current related to left-multiplication is $\jr_\mu$, whose conservation equation
\begin{equation}
    \partial_\mu \jr{}^\mu=0\,,
\end{equation}
is also equivalent to the equations of motion, which yields an equivalent construction of the Lax connection.

\subsection{Root-\texorpdfstring{$\TT$}{TTbar} Deformation}
\label{sec:pcmwz:deform}
It is immediate to write down a deformed Lagrangian, by plugging into eq.~\eqref{eq:deformedlagrangian} the appropriate expressions for the PCM, which we may read off~\eqref{eq:Lpcm}. We have
\begin{equation}
\label{eq:x0x1x2PCM}
    x_0=0,\qquad x_1 = -\tr[j_+ j_-]\,,\qquad
    x_2 =\frac{1}{2}\left(\tr[j_+j_+]\tr[j_-j_-]+\left(\tr[j_+ j_-]\right)^2\right) \, ,
\end{equation}
so that
\begin{equation}
\label{eq:Lpcmgamma}
    L_{\PCM}^{(\gamma)} = \frac{1}{2}\left(
    -\ch(\gamma)\tr[j_+ j_-]+\sh(\gamma)\sqrt{\tr[j_+j_+]\tr[j_-j_-]}
    \right).
\end{equation}
We can equivalently write 
\begin{equation}
    L_{\PCM}^{(\gamma)} = \frac{1}{2}\left(
    -\ch(\gamma)\tr[\jr_+ \jr_-]+\sh(\gamma)\sqrt{\tr[\jr_+\jr_+]\tr[\jr_-\jr_-]}
    \right),
\end{equation}
either by expressing the seed Lagrangian in terms of $\jr_\mu$ in the first place, or by using the explicit form of $j_\mu$ and~$\jr_\mu$.

The above expressions make it manifest that the deformed model still has $G\times G$ symmetry. 
It is possible to derive the resulting conserved current by Noether's theorem. In Appendix~\ref{app:eomPCM} we do so for any Lagrangian which depends on $x_1$ and $x_2$ as in eq.~\eqref{eq:x0x1x2PCM}, $L(x_1,x_2)$, obtaining for the current related to the right-multiplication symmetry
\begin{equation}
\label{eq:deformedPCMeomgeneric}
    \CC{J}_\mu =2\frac{\partial L}{\partial x_1} j_\mu +
    4\frac{\partial L}{\partial x_2} g^{\nu\rho}\tr[j_\mu j_\nu]\,j_\rho\,,\qquad \partial_\mu \CC{J}^\mu=0 \, .
\end{equation}
We immediately see that for the seed Lagrangian $L_0=\tfrac{1}{2}x_1$ this reduces to the current~$j^\mu$.
For the action~\eqref{eq:Lpcmgamma} this gives
\begin{equation}
\label{eq:deformedPCMeom}
    \CC{J}_{\pm}= \ch(\gamma)\,j_\pm - \sh(\gamma) \sqrt{\frac{\tr[j_\pm  j_\pm]}{\tr[j_\mp j_\mp]}}\,j_\mp\,,\qquad
    \partial_{+}\CC{J}_- +\partial_{-}\CC{J}_+=0\,. 
\end{equation}
Note that we may rewrite the Lagrangian~\eqref{eq:Lpcmgamma} as
\begin{equation}
    L_\gamma = -\frac{1}{2}\tr\left[
    j_{+}\CC{J}_-
    \right]=-\frac{1}{2}\tr\left[
    \CC{J}_{+}j_-
    \right]\,.
\end{equation}

The current $\CC{J}_\mu$ is conserved; it is however not flat. To see this, let us consider the commutator of the $\CC{J}_\pm$, which actually obeys a nice relation:
\begin{equation}
\label{eq:commutatorID1}
    [\CC{J}_+,\CC{J}_-]=
    \ch(\gamma)^2 [j_+,j_-]+\sh(\gamma)^2 [j_-,j_+] = [j_+,j_-]\,.
\end{equation}
This relation will be crucial for our construction below. To begin with, it means that
\begin{equation}
    \partial_{+}\CC{J}_{-} - \partial_{-}\CC{J}_{+} +[\CC{J}_+,\CC{J}_-]\neq0\,,
\end{equation}
unless $\gamma=0$, because the first two terms are $\gamma$-dependent while the commutator is not.
However, this structure suggests a modification of the Lax connection which is reminiscent of that discussed in~\cite{Chen:2021aid},
\begin{equation}
\label{eq:pcmLaxdeformed}
    \CC{L}_{\pm}^{(\gamma)}= \frac{j_\pm \pm z\,\CC{J}_{\pm}}{1-z^2}\,.
\end{equation}
By construction, this expression reduces to~\eqref{eq:Laxpcm} when $\gamma=0$ and $\CC{J}_\mu=j_\mu$.
Noting the commutators
\begin{equation}
    [\CC{J}_{\pm},j_{\mp}] = \ch(\gamma)[j_{\pm},j_{\mp}]\,,
\end{equation}
which in particular imply that
\begin{equation}
\label{eq:commutatorID2}
    [\CC{J}_{+},j_{-}]-[j_{+},\CC{J}_{-}]=0\,,
\end{equation}
and exploiting the important relation~\eqref{eq:commutatorID1},
we can work out the condition for $\CC{L}_\mu^{(\gamma)}$ to be flat,%
\footnote{%
We use the shorthands
$A_{[\mu}B_{\nu]}=A_{\mu}B_{\nu}-A_{\nu}B_{\mu}$ and  $A_{(\mu}B_{\nu)}=A_{\mu}B_{\nu}+A_{\nu}B_{\mu}$.
}
\begin{equation}
\label{eq:defPCMflatnessderivation}
\begin{aligned}
0&=\partial_{+}\CC{L}^{(\gamma)}_{-} - \partial_{-}\CC{L}^{(\gamma)}_{+} +[\CC{L}^{(\gamma)}_+,\CC{L}^{(\gamma)}_-]
\\
&=\frac{1}{1-z^2}\left(
\partial_{[+}j_{-]}-z\,\partial_{(+}\mathcal J_{-)}+\frac{[j_+,j_-]-z([j_+,\CC{J}_-]-[\CC{J}_+,j_-])-z^2[\CC{J}_+,\CC{J}_-]}{1-z^2}
\right)\\
&=\frac{1}{1-z^2}\left(
\partial_{[+}j_{-]}+[j_+,j_-]-z\,\partial_{(+}\mathcal J_{-)}
\right)\,.
\end{aligned}
\end{equation}
Using the Maurer-Cartan equation~\eqref{eq:maurercartan} we see that $\CC{L}_\mu^{(\gamma)}$ is flat for any~$z$ if and only if the equations of motion~\eqref{eq:deformedPCMeom} are satisfied.

It is quite clear that we could have performed an analogous construction for the current related to left-multiplication by~$g_0$, which would have given
\begin{equation}
    \widetilde{\CC{J}}_{\pm}= \ch(\gamma)\,\jr_\pm - \sh(\gamma) \sqrt{\frac{\tr[\jr_\pm  \jr_\pm]}{\tr[\jr_\mp \jr_\mp]}}\,\jr_\mp\,,\qquad
    \partial_{+}\widetilde{\CC{J}}_- +\partial_{-}\widetilde{\CC{J}}_+=0\,. 
\end{equation}
This current is related to $\CC{J}_\pm$ much in the same way as $\jr_\pm$ is related to $j_\pm$, namely
\begin{equation}
    \widetilde{\CC{J}}_{\mu} = - \text{Ad}_g \CC{J}_\mu\,.
\end{equation}
In a similar way, we can define the Lax connection in terms of~$\widetilde{\CC{J}}_{\pm}$, namely
\begin{align}\label{alternative_deformed_lax}
    \widetilde{\CC{L}}_{\pm}(z) = \frac{\jr_{\pm} \pm z \widetilde{\CC{J}}_{\pm}}{1 - z^2} \, .
\end{align}
As in the undeformed case, the two choices are related by a gauge transformation of the Lax connection
\begin{align}
    g(\CC{L}_{\pm}(z)+\partial_\pm)g^{-1}=\widetilde{\CC{L}}(z^{-1})\,.
\end{align}

\subsection{Two-Parameter Deformation}
\label{sec:pcmwz:2param}
An important property of the Root-$\TT$ flow is that it commutes with the usual (irrelevant) $\TT$ flow~\cite{Ferko:2022lol}. Following the discussion of Section~\ref{sec:review:doubledef}, for a theory whose ``seed'' Lagrangian is
\begin{equation}
    L_0(x_0,x_1,x_2) = \frac{1}{2}x_1\,,
\end{equation}
we have from eq.~\eqref{eq:doubleformedL}
\begin{equation}
     L_{(\lambda,\gamma)}(x_0,x_1,x_2)
     = \frac{1}{2\lambda}\left(
\sqrt{1+2\lambda x_1^{(\gamma)}+2\lambda^2\left((x_1^{(\gamma)})^2-x_2^{(\gamma)}\right)}-1
    \right),
\end{equation}
with $x_1^{(\gamma)}$ and $x_2^{(\gamma)}$ as in~\eqref{eq:x1x2generic}.
This can be immediately adapted to the case of the PCM by substituting the values~\eqref{eq:x0x1x2PCM} for $x_1$ and $x_2$.
Then, we get a deformed action of the type discussed in Section~\ref{sec:pcmwz:pcm}. In particular, the equations of motion are still given by the conservation of $\CC{J}_\mu$ as in eq.~\eqref{eq:deformedPCMeomgeneric}. We can write $\CC{J}_{\pm}$ a little more explicitly in light-cone coordinates,
\begin{equation}
\label{eq:JmuDoubleDef}
\begin{aligned}
    \CC{J}_{+}&=2\left(\frac{\partial L_{(\lambda,\gamma)}}{\partial x_1}+x_1\frac{\partial L_{(\lambda,\gamma)}}{\partial x_2}\right)j_+ - 2 \frac{\partial L_{(\lambda,\gamma)}}{\partial x_2} \tr[j_+ j_+] j_-\,,\\
    \CC{J}_{-}&=2\left(\frac{\partial L_{(\lambda,\gamma)}}{\partial x_1}+x_1\frac{\partial L_{(\lambda,\gamma)}}{\partial x_2}\right)j_- - 2 \frac{\partial L_{(\lambda,\gamma)}}{\partial x_2} \tr[j_- j_-] j_+\,,
\end{aligned}
\end{equation}
where $j_\mu$ is again the pull-back of the left-invariant Maurer-Cartan one form.%
\footnote{As before, the whole discussion could be repeated in terms of the right-invariant current~$\jr_\mu$.}

We will now see that the Lax connection takes once again the form
\begin{equation}
    \CC{L}_{\pm}^{(\lambda,\gamma)} = \frac{j_\pm \pm z\,\CC{J}_{\pm}}{1-z^2}\,,
\end{equation}
just like in eq.~\eqref{eq:pcmLaxdeformed}, where now $\CC{J}_\mu$ is given by~\eqref{eq:JmuDoubleDef}.
To see why this is the case, we first observe that
\begin{equation}
\begin{aligned}
    \big[\CC{J}_+,j_-\big] =2 \left(\frac{\partial L_{(\lambda,\gamma)}}{\partial x_1}+x_1\frac{\partial L_{(\lambda,\gamma)}}{\partial x_2}\right)\big[j_+,j_-\big],\\
    \big[j_+,\CC{J}_-\big] =2 \left(\frac{\partial L_{(\lambda,\gamma)}}{\partial x_1}+x_1\frac{\partial L_{(\lambda,\gamma)}}{\partial x_2}\right)\big[j_+,j_-\big],
\end{aligned}
\end{equation}
so that in particular
\begin{equation}
    \big[\CC{J}_+,j_-\big]-\big[j_+,\CC{J}_-\big] =0\,,
\end{equation}
like we had before~\eqref{eq:commutatorID2}.
Next, we compute again
\begin{equation}
\label{eq:commutatorcoefficentdoubledef}
    \big[\CC{J}_+,\CC{J}_-\big]=4\left[\left(\frac{\partial L_{(\lambda,\gamma)}}{\partial x_1}+x_1\frac{\partial L_{(\lambda,\gamma)}}{\partial x_2}\right)^2-\left(\frac{\partial L_{(\lambda,\gamma)}}{\partial x_2}\right)^2\big(2x_2 - x_1^2 \big)\right]\,\big[j_+,j_-\big].
\end{equation}
where we used that $\tr[j_+ j_+]\tr[j_- j_-]= 2x_2-x_1^2$ to rewrite the whole coefficient on the right-hand side in terms of $x_1$ and $x_2$.
An explicit computation using the form of~$L_{(\lambda,\gamma)}$ in~\eqref{eq:doubleformedL} then shows that, \textit{for any $\gamma$ and $\lambda$}, the whole coeffiecient on the right-hand side simplifies to one. In other words,
\begin{equation}
    \big[\CC{J}_+,\CC{J}_-\big]=\big[j_+,j_-\big],
\end{equation}
just like in~\eqref{eq:commutatorID1}.  Using these two observations, the derivation of eq.~\eqref{eq:defPCMflatnessderivation} goes through immediately, showing that for this model too the flatness of $\CC{L}_{\pm}^{(\lambda,\gamma)}$ is equivalent to the equations of motion.%
\footnote{%
Conversely, we could obtain a whole class of integrability-preserving Lagrangians $L(x_1,x_2)$ by demanding that the coefficient on the right-hand side of~\eqref{eq:commutatorcoefficentdoubledef} equals one. Unfortunately, it is not immediate to construct the most general solution of this nonlinear partial differential equation.
}
We note that, when $\gamma = 0$, we recover the Lax connection for the $\TT$-deformed PCM which was obtained in \cite{Chen:2021aid}.

\subsection{Adding a Wess-Zumino Term}
\label{sec:pcmwz:wz}
The PCM on the Lie group~$G$ can be supplemented by a Wess-Zumino (WZ) term. The resulting action, prior to any (Root-$\TT$) deformation, reads
\begin{equation}
    \mathcal{S}_{\PCMWZ} = \hay\int\limits_{\partial B}\de^2\sigma L_{\PCM} +\kay\int\limits_B\de^3\sigma\frac{1}{6}\varepsilon^{ijk}\tr\big[j_i\,[j_j,j_k]\big]\,.
\end{equation}
A few remarks are in order. First of all, we have expressed the WZ term as a three-dimensional integral on a space~$B$ whose boundary $\partial B$ is the usual $(1+1)$-dimensional Minkowski space.%
\footnote{%
Here $\varepsilon^{ijk}$ is the three-dimensional Levi-Civita tensor.} 
For this to lead to a well-defined (quantum) theory it may be necessary to impose a constraint for the coefficient $\kay$. In particular, if $G$ is a compact simple Lie group, we should take
\begin{equation}
    4\pi\kay\in\mathbb{Z}\,.
\end{equation}
The coefficient $\hay$, instead, is unconstrained. We could have of course introduced $\hay$ in the discussion of the PCM above, but it would only have amounted to a rescaling of the action.

Let us briefly review the construction of the Lax connection in the undeformed model, see also \textit{e.g.}\ the review \cite{Hoare:2021dix}. First of all $j_\mu$ (and $\jr_\mu$) still satisfy the flatness condition~\eqref{eq:maurercartan}. However, the equations of motion are now modified and read
\begin{equation}
\label{eq:PCMWZeom}
    (\hay+\kay)\partial_{+}j_- + (\hay-\kay)\partial_-j_+= 0\,,
\end{equation}
or equivalently
\begin{equation}
    (\hay-\kay)\partial_{+}\jr_- + (\hay+\kay)\partial_-\jr_+= 0\,.
\end{equation}
It is easy to see that the Lax connection
\begin{equation}
    \CC{L}_\pm = \left(1\mp\frac{\kay}{\hay}\right)\,\frac{j_\pm}{1\mp z}
\end{equation}
reduces to~\eqref{eq:Laxpcm} for $\kay=0$ and that its flatness is equivalent to the equations of motion. We review the computation as it will be useful in a moment. We have
\begin{equation}
\label{eq:flatnessPCMWZ}
\begin{aligned}
    0\,&=\partial_{[+}\CC{L}_{-]}+[\CC{L}_+,\CC{L}_-]\\
    &=\frac{1}{1-z^2}\left(
    (1-z)(1+\tfrac{\kay}{\hay})\partial_+j_- 
    -(1+z)(1-\tfrac{\kay}{\hay})\partial_-j_+
    +(1-\tfrac{\kay^2}{\hay^2})[j_+,j_-]
    \right)\\
    &=\frac{1}{1-z^2}\left(
    \tfrac{\kay}{\hay}-z\right)\left[(1+\tfrac{\kay}{\hay})\partial_+j_- 
    +(1-\tfrac{\kay}{\hay})\partial_-j_+\right],
\end{aligned}
\end{equation}
where in the last step we used the Maurer-Cartan equation to eliminate the commutator. As a result, eq.~\eqref{eq:flatnessPCMWZ} is equivalent to the equations of motion~\eqref{eq:PCMWZeom}.

Let us now consider the case of the Root-$\TT$ deformation (or the combined Root-$\TT$ and $\TT$ deformation) of the PCM with WZ term. The action of the deformed model is simply
\begin{equation}
    \mathcal{S}_{\PCMWZ}^{(\gamma)} = \hay\int\limits_{\partial B}\de^2\sigma L_{\PCM}^{(\gamma)} +\kay\int\limits_B\de^3\sigma\frac{1}{6}\varepsilon^{ijk}\tr\big[j_i\,[j_j,j_k]\big]\,,
\end{equation}
where $L_{\PCM}^{(\gamma)}$ is the very same Lagrangian as in~\eqref{eq:Lpcmgamma}. This construction also applies verbatim if $L_{\PCM}^{(\gamma)}$ is replaced by the doubly-deformed Lagrangian $L_{\PCM}^{(\lambda, \gamma)}$ of (\ref{eq:doubleformedL}). In fact, as we discussed in Section~\ref{sec:review}, the terms of the action which do not couple to the two-dimensional metric do not affect the flow. This is the case for the WZ term, which is topological.
The derivation of the equations of motion splits now in two parts. The part scaling with $\hay$ is exactly as in Section~\ref{sec:pcmwz:deform} (see also Appendix~\ref{app:eomPCM} for the detailed derivation). The part scaling with $\kay$ goes like in the standard WZ term construction. All in all, the equations take the form
\begin{equation}
\label{eq:PCMWZeomlambda}
    \partial_{+}\left(\hay\CC{J}_- +\kay j_-\right)+\partial_{-}\left(\hay\CC{J}_+ -\kay j_+\right)=0\,,
\end{equation}
which should be compared to eq.~\eqref{eq:PCMWZeom} above; a similar equation holds for the right-invariant currents.
These equations of motion suggest a straightforward guess for the Lax connection of the deformed model, namely
\begin{equation}
\label{eq:LaxPCMWZdeformed}
    \CC{L}_{\pm}^{(\gamma)} = \frac{\left(j_\pm \mp\frac{\kay}{\hay}\CC{J}_\pm\right)\pm z\,
    \left(\CC{J}_\pm \mp\frac{\kay}{\hay} j_\pm\right)}{1-z^2}\,.
\end{equation}

Let us now prepare to study the flatness condition for~\eqref{eq:LaxPCMWZdeformed}. By making use of the identities of the previous section, in particular eqs.~\eqref{eq:commutatorID1} and~\eqref{eq:commutatorID2}, we can compute
\begin{equation}
    \big[\CC{L}^{(\gamma)}_+,\CC{L}^{(\gamma)}_-\big]= \frac{1-\frac{\hay^2}{\kay^2}}{1-z^2}\,[j_+,j_-]\,.
\end{equation}
Then the computation follows closely that of eq.~\eqref{eq:flatnessPCMWZ}, and reads 
\begin{equation}
\label{eq:flatnessPCMWZgamma}
\begin{aligned}
    0\,&=\partial_{[+}\CC{L}_{-]}+[\CC{L}_+,\CC{L}_-]\\
    &=\frac{1}{1-z^2}\Big(
    (1-\tfrac{\kay}{\hay}z)\partial_+j_- +(\tfrac{\kay}{\hay}-z)\partial_+ \CC{J}_{-}\\
    &\qquad\qquad\qquad
    -(1-\tfrac{\kay}{\hay}z)\partial_-j_+ +(\tfrac{\kay}{\hay}-z)\partial_- \CC{J}_{+}
    +(1-\tfrac{\kay^2}{\hay^2})[j_+,j_-]
    \Big)\\
    &=\frac{1}{1-z^2}\left(
    \tfrac{\kay}{\hay}-z\right)\left[\partial_+\left(\CC{J}_- +\tfrac{\kay}{\hay}j_-\right) 
    +\partial_-\left(\CC{J}_+ -\tfrac{\kay}{\hay}j_+\right)
    \right],
\end{aligned}
\end{equation}
where again in the last step we used the Maurer-Cartan equation to eliminate the term involving $[j_+,j_-]$. The above equation is therefore equivalent to the equations of motion~\eqref{eq:PCMWZeomlambda}. This construction can be repeated in terms of the right-invariant current too.

\subsection{About the Wess-Zumino-Witten Point}
\label{sec:pcmwz:wzw}
The physics of the PCM with a WZ term depends on the ratio of $\kay/\hay$. In the undeformed model, it is interesting to consider the equations of motion in the limit when $\hay\to\kay$. One finds that the equations of motion imply that the left- and right-invariant currents become chiral,
\begin{equation}
\label{eq:conservationchiral}
    \partial_{+}j_-=0\,,\qquad \partial_{-}\jr_+=0\,.
\end{equation}
In the opposite limit, when $\hay\to-\kay$, the currents $j_\mu$ and $\jr_\mu$ swap roles.
In the quantum theory, these points correspond to the WZW model. At these points, the chiral conservation laws~\eqref{eq:conservationchiral}   lead to the existence of two (chiral and anti-chiral) Ka\v{c}-Moody algebras built out of~$G$.
It is natural to wonder whether this is the case in the deformed model. In the limit $\hay\to\kay$, the equations of motion~\eqref{eq:PCMWZeomlambda} become
\begin{equation}
    \partial_+\left(\CC{J}_- + j_-\right)+\partial_-\left(\CC{J}_+ - j_+\right)=0\,,
\end{equation}
which is not chiral. This is still a classical statement, but it has intriguing implications when we consider a Root-$\TT$ deformation of the WZW model. It points to a rather radical deformation of the current algebra underlying the WZW model, which would be very interesting to study at the quantum level.

\section{Non-Linear Sigma Models}
\label{sec:nlsm}
Let us consider now a sigma model having as target space the quotient $G/H$, where $G$ is a Lie group and $H$ is a subgroup, see also~\cite{Arutyunov:2009ga, Zarembo:2017muf,Seibold:2020ouf} for reviews. Indicating the Lie algebra as $\alg{g}$ and its subalgebra as~$\alg{h}$, we have the vector-space decomposition
\begin{equation}
    \alg{g}=\alg{h}\oplus\alg{f}\,.
\end{equation}
\textit{Symmetric space sigma models} correspond to the special case where the Lie algebra $\alg{g}$ admits a $\mathbb{Z}_2$ automorphism such that%
\footnote{%
The reason for choosing subscript indices $0$ and $2$ rather than $0$ and $1$ will become clear in a few lines, when we will introduce the \textit{semi-symmetric space} sigma model.}
\begin{equation}
\label{eq:SSSMdecomp}
    \alg{g}_0=\alg{h}\,,\qquad
    \alg{g}_2=\alg{f}\,,\qquad
    [\alg{g}_n,\alg{g}_m]\subset\alg{g}_{(n+m)\text{mod}2}\,.
\end{equation}
As we will briefly review below, this fact is sufficient to guarantee the existence of a Lax connection for the model, and we will show that the same is true even if we perform a Root-$\TT$ deformation of the model.

A further generalisation of such cosets is given by \textit{semi-symmetric space sigma models}. These are constructed from Lie supergroups, whose tangent spaces are Lie superalgebras. Let $G$ be a Lie supergroup, and $\mathfrak{g}$ its superalgebra. By definition, the superalgebra has a $\mathbb{Z}_2$ grading, given by the ``Fermion sign'' $(-1)^F$. This fact by itself is not sufficient to guarantee the existence of a Lax connection on a suitably-defined supercoset. If, however, the algebra can be endowed with a $\mathbb{Z}_4$ automorphism~$\Omega$, such that $\Omega^2=(-1)^F$, then it is possible to construct such an integrable supercoset. We decompose the Lie superalgebra in its eigenspaces under $\Omega$,
\begin{equation}
\label{eq:sSSSMdecomp}
    \alg{g}= \alg{g}_0\oplus\alg{g}_1\oplus\alg{g}_2\oplus\alg{g}_3,\qquad
    \Omega\,\alg{g}_n=i^n\,\alg{g}_n\,,\qquad
    [\alg{g}_n,\alg{g}_m]\subset\alg{g}_{(n+m)\text{mod}4},
\end{equation}
where the odd eigenspaces corresponds to fermionic generators and the even ones to bosonic generators. It follows in particular that $\alg{h}=\alg{g}_0$ is a (bosonic) subalgebra of $\alg{g}$. Calling $H$ the corresponding subgroup, the coset space $G/H$ is the target manifold of the semi-symmetric space sigma model. This structure is sufficient to construct a Lax connection for this model, as well as for its Root-$\TT$ deformation.

\subsection{Symmetric-Space Sigma Model (SSSM)}
\label{sec:nlsm:sssm}
Here we consider the coset $G/H$, with $H$ a subgroup of $G$ such that the corresponding Lie algebras enjoy the decomposition~\eqref{eq:SSSMdecomp}.
Similarly to what we did for the PCM in Section~\ref{sec:pcmwz:pcm}, let us consider an element $g\in G$ and denote the pull-back of the (left-invariant) Maurer-Cartan form as $j_\mu = g^{-1}\partial_\mu g$. Since $j_\mu\in\alg{g}=\alg{g}_0\oplus\alg{g}_2$, it can be decomposed as
\begin{equation}
    j_\mu = j^{(0)}_\mu +j^{(2)}_\mu \,,\qquad
    j^{(n)}_\mu\in\alg{g}_n\,.
\end{equation}
Hence $j^{(0)}_\mu$ takes value in the subalgebra $\alg{g}_0=\alg{h}$, while $j^{(2)}_\mu$ takes value in the orthogonal complement. It is worth noting how these two components transform under the subalgebra~$\alg{h}=\alg{g}_0$. Letting $h\in\alg{h}$, and acting on~$g$ by right multiplication, $g'=g\,h$, we see that
\begin{equation}
    (g')^{-1}\partial_\mu g' = \underbrace{h^{-1}j^{(0)}_\mu h +h^{-1}\partial_{\mu}h}_{(j')_\mu^{(0)}\in\alg{g}_0}+\underbrace{h^{-1}j^{(2)}_\mu h}_{(j')_\mu^{(2)}\in\alg{g}_2}\,,
\end{equation}
so that $j^{(0)}_\mu$ transforms like a connection while $j^{(2)}_\mu$ transforms in the adjoint representation of~$\alg{h}$. This also makes it natural to introduce the ``covariant derivative''
\begin{equation}
\label{eq:covariantD}
    D_\mu = \partial_\mu + \left[j^{(0)}_\mu,\, \cdot\,\right]\,,
\end{equation}
as well as the ``field strength''
\begin{equation}
\label{eq:fieldstrength}
    F_{\mu\nu}^{(0)} = \partial_\mu  j^{(0)}_\nu - \partial_\nu  j^{(0)}_\mu + [j^{(0)}_\mu, j^{(0)}_\nu]\,.
\end{equation}

The Lagrangian for the SSSM is then
\begin{equation}
\label{eq:LSSSM}
    L_{\SSSM}= \frac{1}{2}g^{\mu\nu}\tr\left[j^{(2)}_\mu\,j^{(2)}_\nu\right]=-\frac{1}{2}\tr\left[j^{(2)}_+\,j^{(2)}_-\right].
\end{equation}
This is immediately invariant under (local) transformations of~$H$ from the right, as well as under global~$G$ transformations from the left. To see the left $G$-invariance it is sufficient to write
\begin{equation}
    \tr\left[j^{(2)}_\mu\,j^{(2)}_\nu\right]=\tr\left[g^{-1}\partial_\mu g\,P^{(2)}\,g^{-1}\partial_\nu g\right]\,,
\end{equation}
where $P^{(2)}$ projects onto~$\alg{g}_2$. This expression is manifestly invariant under $g\to g'=g_0\,g$.
The equations of motion are equivalent to the covariant conservation of~$j^{(2)}_\mu$
\begin{equation}
\label{eq:eomSSSM}
    g^{\mu\nu}D_\mu\,j^{(2)}_\nu=0\,,
\end{equation}
see Appendix~\ref{app:eomSSSM}.
The appearance of the covariant derivative is natural to ensure symmetry with respect to~$H$.
Note that now $j^{(2)}_\mu$ is (covariantly) conserved but, unlike the case of the PCM, it is not flat --- unlike $j_\mu$, which still satisfies the Maurer-Cartan equation
\begin{equation}
\label{eq:MaurerCartanSSSM}
0=\partial_{[+}j_{-]}+[j_+,j_-]= \underbrace{F_{+-}^{(0)}+\left[j^{(2)}_+,j^{(2)}_-\right]}_{\alg{g}_0}+ \underbrace{D_{+}j^{(2)}_{-}-D_{-}j^{(2)}_{+}}_{\alg{g}_2}\,.
\end{equation}
Note that we decomposed the right-hand side under $\mathbb{Z}_2$, and that the two underbraced terms must separately vanish.

For this model it is  possible to define a Lax connection
\begin{equation}
    \CC{L}_{\pm}= j^{(0)}_\pm + \frac{z\mp1}{z\pm1}j^{(2)}_\pm\,.
\end{equation}
Let us sketch the computation of the flatness condition, in preparation for the case of the Root-$\TT$ deformed model.
We have
\begin{equation}
\begin{aligned}
0\,&=\partial_{[+}\CC{L}_{-]}+\left[\CC{L}_{+},\CC{L}_{-}\right]\\
&=F^{(0)}_{+-} +\left[j^{(2)}_+,j^{(2)}_-\right]+\frac{(z+1)^2D_+j^{(2)}_{-}-(z-1)^2D_-j^{(2)}_{+}}{z^2-1}\,.
\end{aligned}
\end{equation}
We see that the $z$-independent part vanishes due to the $\alg{g}_0$-projection of the Maurer-Cartan equation~\eqref{eq:MaurerCartanSSSM}, while the remaining terms arrange themselves into the $\alg{g}_1$-projection of the Maurer-Cartan equation and the equations of motion~\eqref{eq:eomSSSM}, which is precisely what we~need.

\subsection{Root-\texorpdfstring{$\TT$}{TTbar} deformation of the SSSM}
\label{sec:nlsm:sssmdef}
The SSSM Lagrangian~\eqref{eq:LSSSM} is again of the form $L(x_0,x_1,x_2)$ (see Section~\ref{sec:review}) with $x_0=0$ and
\begin{equation}
\label{eq:x1x2SSSM}
    x_1 = -\tr[j^{(2)}_+ j^{(2)}_-]\,,\qquad
    x_2 =\frac{1}{2}\left(\tr[j^{(2)}_+j^{(2)}_+]\tr[j^{(2)}_-j^{(2)}_-]+\left(\tr[j^{(2)}_+ j^{(2)}_-]\right)^2\right) \,,
\end{equation}
which is the same form as~\eqref{eq:x0x1x2PCM}.
As a result, the deformed Lagrangian $L_{\SSSM}^{(\gamma)}$ takes the same functional form as in~\eqref{eq:Lpcmgamma}, that is
\begin{equation}
\label{eq:LSSSMgamma}
    L_{\SSSM}^{(\gamma)} = \frac{1}{2}\left(
    -\ch(\gamma)\tr[j^{(2)}_+ j^{(2)}_-]+\sh(\gamma)\sqrt{\tr[j^{(2)}_+j^{(2)}_+]\tr[j^{(2)}_-j^{(2)}_-]}
    \right).
\end{equation}
This Lagrangian is manifestly invariant under local~$H$ transformations as well as under global~$G$ transformations from the left $g'=g_0g$, with $g_0$ a constant element of~$G$.
The equations of motion can be derived as in Appendix~\ref{app:eomSSSM}, and they read
\begin{equation}
\label{eq:conservationSSMdeform}
    D_\mu \CC{J}_{(2)}^\mu=0\,,\qquad
    \CC{J}^{(2)}_\mu = 2\frac{\partial L}{\partial x_1}j^{(2)}_\mu+4g^{\nu\rho}\frac{\partial L}{\partial x_2}\tr[j^{(2)}_\mu j^{(2)}_\nu]\,j^{(2)}_\rho\,.
\end{equation}
By analogy with the discussion of Section~\ref{sec:pcmwz:deform}, it is natural to define the Lax connection
\begin{equation}
    \CC{L}_{\pm}^{(\gamma)} = j^{(0)}_\pm + \frac{(z^2+1)\,j^{(2)}_\pm \pm 2z\,\CC{J}_\pm^{(2)}}{z^2-1}\,,
\end{equation}
which reduces to $\CC{L}_{\pm}$ in the undeformed model where $\CC{J}_\pm^{(2)}=j_\pm^{(2)}$.

To check the flatness condition, let us start by evaluating more explicitly $\CC{J}^{(2)}_\mu$. This takes the same form as in~\eqref{eq:JmuDoubleDef},
\begin{equation}
\label{eq:JmuSSSM}
\begin{aligned}
    \CC{J}^{(2)}_{+}&=2\left(\frac{\partial L_{\SSSM}^{(\gamma)}}{\partial x_1}+x_1\frac{\partial L_{\SSSM}^{(\gamma)}}{\partial x_2}\right)j^{(2)}_+ - 2 \frac{\partial L_{\SSSM}^{(\gamma)}}{\partial x_2} \tr[j^{(2)}_+ j^{(2)}_+] j^{(2)}_-\,,\\
    \CC{J}^{(2)}_{-}&=2\left(\frac{\partial L_{\SSSM}^{(\gamma)}}{\partial x_1}+x_1\frac{\partial L_{\SSSM}^{(\gamma)}}{\partial x_2}\right)j^{(2)}_- - 2 \frac{\partial L_{\SSSM}^{(\gamma)}}{\partial x_2} \tr[j^{(2)}_- j^{(2)}_-] j^{(2)}_+\,.
\end{aligned}
\end{equation}
Like in the case of PCM, we note the analogue of the identities \eqref{eq:commutatorID1} and \eqref{eq:commutatorID2}, namley
\begin{equation}
    \big[\CC{J}^{(2)}_\pm,j^{(2)}_\mp\big] =2 \left(\frac{\partial L_{\SSSM}^{(\gamma)}}{\partial x_1}+x_1\frac{\partial L_{\SSSM}^{(\gamma)}}{\partial x_2}\right)\big[j^{(2)}_\pm,j^{(2)}_\mp\big],
\end{equation}
so that in particular
\begin{equation}
\label{eq:commutatorid1sssm}
    \big[\CC{J}^{(2)}_+,j^{(2)}_-\big]-\big[j^{(2)}_+,\CC{J}^{(2)}_-\big] =0\,.
\end{equation}
Additionally
\begin{equation} 
\label{eq:commutatorid2sssm}
    \big[\CC{J}^{(2)}_+,\CC{J}^{(2)}_-\big]=\big[j^{(2)}_+,j^{(2)}_-\big]\,,
\end{equation}
which can be checked by using that $L_{\SSSM}^{(\gamma)}$ obeys the differential equation
\begin{equation}
    4\left[\left(\frac{\partial L_{\SSSM}^{(\gamma)}}{\partial x_1}+x_1\frac{\partial L_{\SSSM}^{(\gamma)}}{\partial x_2}\right)^2-\left(\frac{\partial L_{\SSSM}^{(\gamma)}}{\partial x_2}\right)^2\big(2x_2-(x_1)^2\big)\right]=1\,,
\end{equation}
exactly as in~\eqref{eq:commutatorcoefficentdoubledef}. The same differential equation is satisfied by the doubly-deformed Lagrangian $L_{SSSM}^{(\lambda, \gamma)}$ of (\ref{eq:doubleformedL}), so this entire argument can be repeated for the $\TT$ and Root-$\TT$ deformed SSSM, defining the currents (\ref{eq:JmuSSSM}) appropriately (see Section~\ref{sec:nlsm:sSSSMWZ} below). Using these observations we find that
\begin{equation}
\begin{aligned}
0\,&=\partial_{[+}\CC{L}_{-]}+\left[\CC{L}_{+},\CC{L}_{-}\right]\\
&=F^{(0)}_{+-} +\left[j^{(2)}_+,j^{(2)}_-\right]+\frac{z^2+1}{z^2-1}\left(D_+j^{(2)}_- -D_-j^{(2)}_+\right)+
\frac{2z}{z^2-1}\left(D_+j^{(2)}_- +D_-j^{(2)}_+\right)\,,
\end{aligned}
\end{equation}
where the various pieces reproduce the Maurer-Cartan equation (more precisely, its $\alg{g}_0$ and $\alg{g}_2$ components), and the equations of motion.

\subsection{Semi-Symmetric-Space Sigma Model (sSSSM)}
\label{sec:nlsm:ssssm}
In the sSSSM, $G$ is a supergroup with superalgebra $\alg{g}$, and it enjoys a $\mathbb{Z}_4$ decomposition of the form~\eqref{eq:sSSSMdecomp}. We consider the coset $G/H$ with $\alg{h}=\alg{g}_0$. Accordingly, we decompose the pull-back of the Maurer-Cartan form as
\begin{equation}
    j_\mu = j_\mu^{(0)}+ j_\mu^{(1)}+ j_\mu^{(2)}+ j_\mu^{(3)}\,,\qquad
    j_\mu^{(n)}\in\alg{g}_n.
\end{equation}
The Maurer-Cartan equation itself can be decomposed as
\begin{equation}
    \alg{g}_n:\qquad
    0=\partial_{+}j^{(n)}_{-}-\partial_{-}j^{(n)}_{+} + \sum_{j=0}^3\left[j_+^{(j)},j_-^{(n-j)}\right]\,.
\end{equation}
for $n=0,\dots,3$, where the indices of the current are understood to take values modulo~4.
It is convenient to highlight the special role played by $\alg{h}=\alg{g}_0$ by expressing the Maurer-Cartan equation in terms of the covariant derivative~\eqref{eq:covariantD} and field strength~\eqref{eq:fieldstrength},%
\footnote{Recall that in our conventions $\varepsilon^{+-}=-\varepsilon^{-+}=\tfrac{1}{2}$, while $g^{+-}=g^{-+}=-\tfrac{1}{2}$ and $g_{+-}=g_{-+}=-2$.}
\begin{equation}
\label{eq:sSSSMMaurerCartan}
\begin{aligned}
    \alg{g}_0:&\qquad0=F_{+-}^{(0)}+\left[j_+^{(2)},j_-^{(2)}\right]+2\varepsilon^{\mu\nu}\left[j_\mu^{(1)},j_\nu^{(3)}\right],\\
    \alg{g}_1:&\qquad0=D_+j^{(1)}_- - D_-j^{(1)}_+ + 2\varepsilon^{\mu\nu}\left[j_\mu^{(2)},j_\nu^{(3)}\right],\\
    \alg{g}_2:&\qquad0=D_+j^{(2)}_- - D_-j^{(2)}_++\varepsilon^{\mu\nu}\left(\left[j_\mu^{(1)},j_\nu^{(1)}\right]+\left[j_\mu^{(3)},j_\nu^{(3)}\right]\right),\\
    \alg{g}_3:&\qquad0=D_+j^{(3)}_- - D_-j^{(3)}_+ + 2\varepsilon^{\mu\nu}\left[j_\mu^{(2)},j_\nu^{(1)}\right].
\end{aligned}
\end{equation}

The Lagrangian of the sSSSM can be written as~\cite{Metsaev:1998it}%
\footnote{%
The two terms in the sum are separately~$\mathbb{Z}_4$-invariant. The model is only integrable if their relative coefficient is~$\pm1$; this is also the only choice that guarantees $\kappa$-symmetry when one wants to interpret this models as a string theory~\cite{Arutyunov:2009ga}.
}
\begin{equation}
    L_{\sSSSM} = \frac{1}{2}\text{str}\left[g^{\mu\nu}j^{(2)}_{\mu}j^{(2)}_{\nu}+\varepsilon^{\mu\nu}j^{(1)}_{\mu}j^{(3)}_{\nu}\right],
\end{equation}
which is written in terms of the supertrace. The equations of motion then take the form (see Appendix~\ref{app:eomsSSSM})
\begin{equation}
\begin{aligned}
    \alg{g}_1:&\qquad
    \varepsilon^{\mu\nu}D_{\mu}j^{(1)}_\nu = \left(\varepsilon^{\mu\nu}-2g^{\mu\nu}\right)\left[j^{(2)}_\mu,j^{(3)}_\nu\right],\\
    \alg{g}_2:&\qquad
    g^{\mu\nu}D_{\mu}j^{(2)}_\nu =\frac{1}{2}\varepsilon^{\mu\nu}\left(\left[j^{(1)}_\mu,j^{(1)}_\nu\right]-\left[j^{(3)}_\mu,j^{(3)}_\nu\right]\right),\\
    \alg{g}_3:&\qquad
    \varepsilon^{\mu\nu}D_{\mu}j^{(3)}_\nu = \left(\varepsilon^{\mu\nu}+2g^{\mu\nu}\right)\left[j^{(2)}_\mu,j^{(1)}_\nu\right].
\end{aligned}
\end{equation}
These expressions can further be simplified by using~\eqref{eq:sSSSMMaurerCartan}, giving
\begin{equation}
\label{eq:eomsSSSMreduced}
\begin{aligned}
    \alg{g}_1:&&\qquad
    0=& \left(\varepsilon^{\mu\nu}-g^{\mu\nu}\right)\left[j^{(2)}_\mu,j^{(3)}_\nu\right],\\
    \alg{g}_2:&&\qquad
    g^{\mu\nu}D_{\mu}j^{(2)}_\nu =&\, \frac{1}{2}\varepsilon^{\mu\nu}\left(\left[j^{(1)}_\mu,j^{(1)}_\nu\right]-\left[j^{(3)}_\mu,j^{(3)}_\nu\right]\right),\\
    \alg{g}_3:&&\qquad
    0 =& \left(\varepsilon^{\mu\nu}+g^{\mu\nu}\right)\left[j^{(2)}_\mu,j^{(1)}_\nu\right].
\end{aligned}
\end{equation}

The Lax connection is~\cite{Bena:2003wd}
\begin{equation}
    \CC{L}_\pm=
    j^{(0)}_\pm +\sqrt{\frac{z+1}{z-1}}\,j^{(1)}_{\pm}+\frac{z\mp1}{z\pm1}\,j^{(2)}_{\pm} +\sqrt{\frac{z-1}{z+1}}\,j^{(3)}_{\pm}\,.
\end{equation}
The flatness condition for $\CC{L}_\pm$ can then be expressed over the components~$\alg{g}_n$, yielding:
\begin{equation}
\begin{aligned}
    \alg{g}_0:&\qquad0=F_{+-}^{(0)}+\left[j_+^{(2)},j_-^{(2)}\right]+2\varepsilon^{\mu\nu}\left[j_\mu^{(1)},j_\nu^{(3)}\right],\\
    \alg{g}_1:&\qquad0=D_+j^{(1)}_- - D_-j^{(1)}_+ + \frac{(z-1)^2}{(z+1)^2}\left[j_+^{(2)},j_-^{(3)}\right]+\left[j_+^{(3)},j_-^{(2)}\right],\\
    \alg{g}_2:&\qquad0=\frac{z+1}{z-1}\left(D_+j^{(2)}_- + \left[j_+^{(1)},j_-^{(1)}\right]\right) -
    \frac{z-1}{z+1}\left(D_-j^{(2)}_+ - \left[j_+^{(3)},j_-^{(3)}\right]\right),\\
    \alg{g}_3:&\qquad0=D_+j^{(3)}_- - D_-j^{(3)}_+ + \left[j_+^{(2)},j_-^{(1)}\right]+\frac{(z+1)^2}{(z-1)^2}\left[j_+^{(1)},j_-^{(2)}\right].
\end{aligned}
\end{equation}
We see that the $\alg{g}_0$ component matches with the 0-th component of the Maurer-Cartan equation. The $\alg{g}_1$ component, evaluted at $z=0$, gives the corresponding component of the Maurer-Cartan equation, while its residue at the double-pole gives the equations of motion~\eqref{eq:eomsSSSMreduced}; similarly for the $\alg{g}_3$ component. Finally, evaluating the $\alg{g}_2$ component at $z=0$ and $z=i$ gives the Maurer-Cartan and equations of motion, respectively.

\subsection{Root-\texorpdfstring{$\TT$}{TTbar} deformation of the sSSSM}
\label{sec:nlsm:ssssmdef}
The action of the sSSSM is still of the form $L(x_0,x_1,x_2)$. Moreover,  $x_1$ and $x_2$ take the same form as in the (bosonic) SSSM if we express them in terms of~$j^{(2)}_\mu$ up to exchaging the trace for the supertrace,
\begin{equation}
    x_1 = -\text{str}\left[j^{(2)}_+ j^{(2)}_-\right]\,,\qquad
    x_2 =\frac{1}{2}\left(\text{str}\left[j^{(2)}_+j^{(2)}_+\right]\text{str}\left[j^{(2)}_-j^{(2)}_-\right]+\left(\text{str}\left[j^{(2)}_+ j^{(2)}_-\right]\right)^2\right) \,,
\end{equation}
see also eq.~\eqref{eq:x1x2SSSM}. The main new ingredient is that here $x_0\neq0$, and instead
\begin{equation}
    x_0=\frac{1}{2}\varepsilon^{\mu\nu}\text{str}\left[j^{(1)}_{\mu}\,j^{(3)}_{\nu}\right].
\end{equation}
Still, the deformed Lagrangian~$L_{\sSSSM}^{(\gamma)}$ can be easily found from~\eqref{eq:deformedlagrangian}, and the doubly-deformed version $L_{\sSSSM}^{(\lambda, \gamma)}$ from (\ref{eq:doubleformedL}). The equations of motion of the deformed model can be expressed in terms of
\begin{equation}
    \CC{J}^{(2)}_\mu=2\frac{\partial L_{\sSSSM}^{(\gamma)}}{\partial x_1}j^{(2)}_\mu +4\frac{\partial L_{\sSSSM}^{(\gamma)}}{\partial x_2} \,g^{\nu\rho}\text{str}\left[j^{(2)}_\mu j^{(2)}_\nu\right]j^{(2)}_\rho \,,
\end{equation}
which generalises the expression in eq.~\eqref{eq:conservationSSMdeform}.  Note however that the form of the equations of motion is rather more complicated now, due to the contributions from the odd currents in~$x_0$. We refer the reader to appendix~\ref{app:eomsSSSM} for their derivation. They are
\begin{equation}
\label{eq:eomsSSSMreduceddef}
\begin{aligned}
    \alg{g}_1:&&\qquad
    0=& \left[\left(\varepsilon^{\mu\nu}j^{(2)}_\mu-g^{\mu\nu}\CC{J}^{(2)}_\mu\right),j^{(3)}_\nu\right],\\
    \alg{g}_2:&&\qquad
    g^{\mu\nu}D_{\mu}\CC{J}^{(2)}_\nu =&\, \frac{1}{2} \varepsilon^{\mu\nu}\left(\left[j^{(1)}_\mu,j^{(1)}_\nu\right]-\left[j^{(3)}_\mu,j^{(3)}_\nu\right]\right),\\
    \alg{g}_3:&&\qquad
    0 =& \left[\left(\varepsilon^{\mu\nu}j^{(2)}_\mu+g^{\mu\nu}\CC{J}^{(2)}_\mu\right),j^{(1)}_\nu\right],
\end{aligned}
\end{equation}
and it is immediate to see that they reduce to~\eqref{eq:eomsSSSMreduced} in the undeformed limit when~$\CC{J}^{(2)}_\mu=j^{(2)}_\mu$. We now define the Lax connection
\begin{equation}
    \CC{L}_\pm^{(\gamma)}=
    j^{(0)}_\pm +\sqrt{\frac{z+1}{z-1}}\,j^{(1)}_{\pm}+\frac{\left(z^2+1\right)\,j^{(2)}_\pm \mp 2z\,\CC{J}^{(2)}_{\pm}}{z^2-1}+\sqrt{\frac{z-1}{z+1}}\,j^{(3)}_{\pm}\,.
\end{equation}
Clearly the manipulations involving terms which do not involve $j^{(1)}_\mu$ or $j^{(3)}_\mu$ go through exactly like for the SSSM. In particular, we can still use equations~\eqref{eq:commutatorid1sssm}--\eqref{eq:commutatorid2sssm}.
The flatness condition for $\CC{L}_\pm^{(\gamma)}$ then becomes%
\footnote{%
Recall that in our conventions $A_{[\mu}B_{\nu]}=A_{\mu}B_{\nu}-A_{\nu}B_{\nu}$.
}
\begin{equation}
\begin{aligned}
    \alg{g}_0:&\ 0=F_{+-}^{(0)}+\left[j_+^{(2)},j_-^{(2)}\right]+2\varepsilon^{\mu\nu}\left[j_\mu^{(1)},j_\nu^{(3)}\right],\\[0.2cm]
    \alg{g}_1:&\ 0=D^{\phantom{|}}_{[+}j^{(1)}_{-]}  + \left[\frac{(z^2+1)j_+^{(2)}-2z\CC{J}_+^{(2)}}{(z+1)^2},j_-^{(3)}\right]+\left[j_+^{(3)},\frac{(z^2+1)j_+^{(2)}+2z\CC{J}_-^{(2)}}{(z+1)^2}\right],\\[0.2cm]
    \alg{g}_2:&\ 0=
    \frac{z^2+1}{z^2-1}D^{\phantom{|}}_{[+}j^{(2)}_{-]}
    +\frac{2z}{z^2-1}D^{\phantom{|}}_{(+}\CC{J}^{(2)}_{-)}
    + \frac{z+1}{z-1}\left[j_+^{(1)},j_-^{(1)}\right] +
    \frac{z-1}{z+1}\left[j_+^{(3)},j_-^{(3)}\right],\\[0.2cm]
    \alg{g}_3:&\ 0=D^{\phantom{|}}_{[+}j^{(3)}_{-]} + \left[\frac{(z^2+1)j_+^{(2)}-2z\CC{J}_+^{(2)}}{(z-1)^2},j_-^{(3)}\right]+\left[j_+^{(3)},\frac{(z^2+1)j_-^{(2)}+2z\CC{J}_-^{(2)}}{(z-1)^2}\right].
\end{aligned}
\end{equation}
We now require these equations, in particular the last three, to hold for any~$z\in\mathbb{C}$.
We can see that the $\alg{g}_1$ and $\alg{g}_3$ components give the components of the Maurer-Cartan equations at $z=0$, and give the equations of motion when taking the residues of the double poles. The component along  $\alg{g}_2$ gives the Maurer-Cartan equation for~$z=0$ and the equations of motion for $z=\pm i$. This shows the equivalence of the flatness condition for $\CC{L}_\pm^{(\gamma)}$ with the equations of motion.

\subsection{sSSSM with Wess-Zumino Term}
\label{sec:nlsm:sSSSMWZ}

For certain semi-symmetric superspaces, it is possible to add a non-trivial Wess-Zumino coupling while preserving integrability \cite{Cagnazzo:2012se}. As before, let $G$ be a supergroup with Lie superalgebra $\mathfrak{g}$. If $G$ is a simple supergroup, the candidate Wess-Zumino term vanishes identically, so we will not consider models of this form in the current section. However, a Wess-Zumino coupling exists and is non-trivial if $G$ is a direct product $G = G' \times G'$ where $G'$ is another supergroup. For the present discussion, we will assume that $G$ admits such a direct product structure. In particular, this means that $\alg{g} = \alg{g}' \oplus \alg{g}'$, and such a Lie algebra always admits a $\mathbb{Z}_4$ action given by the semi-graded permutation of the two factors. This choice of grading goes under the name of permutation cosets, see~\cite{Cagnazzo:2012se} for further details. We use this structure to decompose $\alg{g} = \alg{g}_0 \oplus \alg{g}_1 \oplus \alg{g}_2 \oplus \alg{g}_3$ as in equation (\ref{eq:sSSSMdecomp}). We again decompose the current $j_\mu$ using the grading of $\alg{g}$ as
\begin{equation}
    j_\mu = j_\mu^{(0)} + j_\mu^{(1)} + j_\mu^{(2)} + j_\mu^{(3)} \, , \qquad j_\mu^{(n)} \in \alg{g}_n \, .
\end{equation}
In terms of the components in this decomposition, the action for the semi-symmetric space sigma model with Wess-Zumino term takes the form~\cite{Cagnazzo:2012se}
\begin{equation}
\label{eq:semi_plus_wz}
\begin{aligned}
    \mathcal{S}_{\text{SMWZ}} &= \hay\int\limits_{\partial B}\de^2\sigma \frac{1}{2}\str \left( g^{\mu \nu} j_\mu^{(2)} j_\nu^{(2)} + \ell \epsilon^{\mu \nu} j_\mu^{(1)} j_\nu^{(3)} \right) \\
    &\qquad + \kay\int\limits_{B}\de^3\sigma \frac{1}{3}\varepsilon^{ijk}  \str \left( j^{(2)}_i [ j^{(2)}_j , j^{(2)}_k ] + 3 j^{(1)}_i [ j^{(3)}_j , j^{(2)}_k ] \right) \, .
\end{aligned}
\end{equation}
A few comments about this action are in order. The first is that, although the current $j_\mu$ itself admits a decomposition according to the grading of $\alg{g}$, the supertrace is not compatible with this new $\mathbb{Z}_4$ grading. In particular, 
one may have a non-vanishing supertrace $\str (\mathfrak g_m\mathfrak g_n)$ not only if\footnote{In Section~
\ref{sec:nlsm:ssssm}, for simplicity,  when deriving the equations of motion we were implicitly assuming compatibility of the supertrace with the $\mathbb{Z}_4$ grading, i.e.~$\str (\mathfrak g_m\mathfrak g_n)$ only if $(m+n) \text{ mod } 4=0$. See also appendix~\ref{app:eomsSSSM}. However, it is worth stressing that the final expression for the equations of motion is the same even when relaxing this assumption, thanks to the fact that the Lagrangian is given by the supertrace of an element of grading 0.} $(m+n) \text{ mod } 4=0$ but also if $(m+n) \text{ mod } 4=2$. This is in fact reflected in  the Wess-Zumino term on the second line of (\ref{eq:semi_plus_wz}), which \emph{does not} respect the $\mathbb{Z}_4$ grading. In particular, this means that the equations of motion for this theory will mix terms from different subspaces $\alg{g}_n$, so that we can no longer decompose equations according to the grading and demand that the components in each subspace vanish separately. However, it is important to remark that the the second line of \eqref{eq:semi_plus_wz} is chosen so that it gives rise to a well-defined WZ term, whose variation reduces to a 2-dimensional integral.

The second comment is that we have introduced an additional parameter $\ell$ which controls the relative scaling between the sigma model term and the odd-current term. Taking $\kay \to 0$ and $\ell \to 1$ recovers the sSSSM with the usual conventions, up to the overall scaling of the action controlled by $\hay$. Although we fixed $\ell = 1$ in the preceding discussion, we will see that it is not consistent to do so when we include the Wess-Zumino term, and in fact the parameters are required to satisfy a constraint $\ell^2 = 1 - \frac{\kay^2}{\hay^2}$ in order to maintain integrability.%
\footnote{%
In the case where the sigma model should be interpreted as a string theory, this requirement can be understood as enforcing $\kappa$-symmetry~\cite{Cagnazzo:2012se}.
}

The equations of motion arising from \eqref{eq:semi_plus_wz} can be written as
\begin{equation}
\label{eq:eom_semi_plus_wz}
    \begin{aligned}
    0 &= \left( \ell \varepsilon^{\mu\nu}-g^{\mu\nu}\right)\left[j^{(2)}_\mu,j^{(3)}_\nu\right] - \frac{\kay}{\hay} \varepsilon^{\mu \nu} [ j^{(2)}_\mu , j^{(1)}_\nu ]  ,  \\
    g^{\mu\nu}D_{\mu}j^{(2)}_\nu &=  \frac{\ell}{2} \varepsilon^{\mu\nu} \left(   \left[j^{(1)}_\mu,j^{(1)}_\nu\right] -  \left[j^{(3)}_\mu,j^{(3)}_\nu\right] \right)+ \frac{\kay}{\hay}\varepsilon^{\mu\nu}   \left( [ j^{(1)}_\mu , j^{(3)}_\nu ] +  [ j^{(2)}_\mu , j^{(2)}_\nu ] \right) , \\
    0 &= \left( \ell \varepsilon^{\mu\nu}+g^{\mu\nu}\right)\left[j^{(2)}_\mu,j^{(1)}_\nu\right] + \frac{\kay}{\hay} \varepsilon^{\mu \nu} [ j_\mu^{(2)} , j_\nu^{(3)} ] ,
\end{aligned}
\end{equation}
and we can see that the terms proportional to $\frac{\kay}{\hay}$ explicitly break the grading. However, the Maurer-Cartan identity for $j_\mu$ is unchanged, so we may still use the same equations (\ref{eq:sSSSMMaurerCartan}) as in discussion of the sSSSM without WZ term.

The Lax connection for this model takes the form~\cite{Cagnazzo:2012se}
\begin{equation}
\label{eq:undeformed_semi_wz}
\begin{aligned}
    \CC{L}_{\pm} &= j_{\pm}^{(0)} + \ell \frac{z^2 + 1}{z^2 - 1} j_{\pm}^{(2)} \pm \left( \frac{\kay}{\hay} - \frac{2 \ell z}{z^2 - 1 } \right) j_{\pm}^{(2)} + \left( z + \frac{\ell}{1 - \frac{\kay}{\hay} } \right) \sqrt{ \frac{\ell \left( 1 - \frac{\kay}{\hay} \right) }{z^2 - 1 } } j_{\pm}^{(1)}  \\
    &\quad + \left( z - \frac{\ell}{1 + \frac{\kay}{\hay} } \right) \sqrt{ \frac{ \ell \left(  1 + \frac{\kay}{\hay} \right) }{ z^2 - 1 } } j_{\pm}^{(3)} \, .
\end{aligned}
\end{equation}
Although the equations of motion (and thus the flatness condition for the Lax connection) do not respect the $\mathbb{Z}_4$ grading of $\CC{g}$, they of course still respect the $\mathbb{Z}_2$ grading of the algebra into bosonic and fermionic subalgebras $\CC{g}_{\text{B}}$ and $\CC{g}_{\text{F}}$. Projecting the equation $\partial_+ \CC{L}_- - \partial_- \CC{L}_+ + [ \CC{L}_+ , \CC{L}_- ] = 0$ onto these subspaces gives the equations
\begin{equation}
\label{eq:bosonic_flatness_lax}
\begin{aligned}
    \CC{g}_{\text{B}} : \; 0 &= F_{+-}^{(0)} + \left( \ell \frac{z-1}{z+1} + \frac{\kay}{\hay} \right) \left( \ell \frac{z+1}{z-1} - \frac{\kay}{\hay} \right) [ j^{(2)}_+ , j^{(2)}_- ]    \\
    &\  + \left( \left( z + \frac{\ell}{1 - \frac{\kay}{\hay} } \right) \left( z - \frac{\ell}{1 + \frac{\kay}{\hay} } \right) \frac{ \ell \sqrt{ 1 - \frac{\kay^2}{\hay^2} } }{z^2 - 1} \right) \left( [ j_+^{(1)} , j_-^{(3)} ] + [ j_+^{(3)} , j_-^{(1)} ]  \right)  \\
    &\   + \left( \ell \frac{z + 1}{z - 1} -  \frac{\kay}{\hay} \right) D_+ j_-^{(2)} - \left( \ell \frac{z - 1}{z + 1} + \frac{\kay}{\hay} \right) D_- j_+^{(2)}  \\
    &\  + \left( z + \frac{\ell}{1 - \frac{\kay}{\hay} } \right)^2  \frac{ \ell \left( 1 - \frac{\kay}{\hay} \right) }{z^2 - 1 }  [ j_+^{(1)} , j_-^{(1)} ]  + \left( z - \frac{\ell}{1 + \frac{\kay}{\hay} } \right)^2  \frac{ \ell \left( 1 + \frac{\kay}{\hay} \right) }{ z^2 - 1 }  [ j_+^{(3)} , j_-^{(3)} ]  \, .
\end{aligned}
\end{equation}
and
\begin{equation}
\label{eq:fermionic_flatness_lax}
\begin{aligned}
    \CC{g}_{\text{F}} : \; 0 &= \left( z + \frac{\ell}{1 - \frac{\kay}{\hay} } \right) \sqrt{\ell \left( 1 - \frac{\kay}{\hay} \right) } \left( D_+ j_-^{(1)} - D_- j_+^{(1)} \right)  \\
    &\ + \left( \ell \frac{z - 1}{z + 1} + \frac{\kay}{\hay} \right) \left( z - \frac{\ell}{1 + \frac{\kay}{\hay} } \right) \sqrt{ \ell \left( 1 + \frac{\kay}{\hay} \right) } [ j_+^{(2)} , j_-^{(3)} ]  \\
    &\ + \left( \ell \frac{z + 1}{z - 1} -  \frac{\kay}{\hay} \right) \left( z - \frac{\ell}{1 + \frac{\kay}{\hay} } \right) \sqrt{ \ell \left( 1 + \frac{\kay}{\hay} \right) }  [ j_+^{(3)} , j_-^{(2)} ]  \\
    &\quad + \left( z - \frac{\ell}{1 + \frac{\kay}{\hay} } \right) \sqrt{ \ell \left( 1 + \frac{\kay}{\hay} \right) } \left( D_+ j_-^{(3)} - D_- j_+^{(3)} \right)  \\
    &\quad + \left( \ell \frac{z - 1}{z + 1} + \frac{\kay}{\hay} \right) \left( z + \frac{\ell}{1 - \frac{\kay}{\hay} } \right) \sqrt{ \ell \left( 1 - \frac{\kay}{\hay} \right)  } [ j_+^{(2)} , j_-^{(1)} ]  \\
    &\quad +  \left( \ell \frac{z + 1}{z - 1} -  \frac{\kay}{\hay} \right) \left( z + \frac{\ell}{1 - \frac{\kay}{\hay} } \right) \sqrt{ \ell \left( 1 - \frac{\kay}{\hay} \right) } [ j_+^{(1)} , j_-^{(2)} ] \, .
\end{aligned}
\end{equation}
By substituting the equations of motion \eqref{eq:eom_semi_plus_wz}, as well as the Maurer-Cartan identities \eqref{eq:sSSSMMaurerCartan} into the right sides of equations \eqref{eq:bosonic_flatness_lax} and \eqref{eq:fermionic_flatness_lax}, one finds that the resulting expressions vanish identically if and only if
\begin{align}\label{eq:ell_constraint}
    \ell^2 = 1 - \frac{\kay^2}{\hay^2} \, , 
\end{align}
as we mentioned above. When the parameters are related in this way, the equations of motion and flatness of $j_\mu$ imply the flatness of the Lax connection for any value of $z$. The converse is also true, as one can show by evaluating the flatness conditions at several values of $z$ to obtain several independent equations and then solving the resulting system. For instance, one can obtain three independent equations from evaluating the bosonic projection (\ref{eq:bosonic_flatness_lax}) at $z=0$ and extracting the two residues at $z = 1$ and at $z = -1$. These three equations can be solved simultaneously to yield expressions for $F_{+-}$, $D_+ j^{(2)}_-$, and $D_- j_+^{(2)}$ which are equivalent to the $\mathfrak{g}_0$ and $\mathfrak{g}_2$ parts of the Maurer-Cartan identity, along with the equation of motion for $j_\mu^{(2)}$. Similarly, one can extract four independent equations from the fermionic projection \eqref{eq:fermionic_flatness_lax} by computing the two residues at $z = \pm 1$ and then evaluating the equation at two other values of $z$, for instance $z=0$ and $z = \ell/(1 + \frac{\kay}{\hay})$. The resulting system of four equations is equivalent to the $\alg{g}_1$ and $\alg{g}_3$ components of the Maurer-Cartan identity, along with the two equations of motion for $j_\mu^{(1)}$ and $j_\mu^{(3)}$ on the first and third lines of \eqref{eq:eom_semi_plus_wz}. This shows that the flatness of the Lax connection for any $z$ implies the equations of motion and the Maurer-Cartan identities, which completes the second part of the equivalence.

\subsection{Two-Parameter Deformation of the sSSSM with WZ Term}
\label{sec:nlsm:2param}

We now consider a two-parameter family of models which arise from deforming the semi-symmetric space sigma model action \eqref{eq:eom_semi_plus_wz} by both $\TT$ and Root-$\TT$. Our Lagrangian $L_{(\lambda, \gamma)}$ satisfies the two simultaneous flow equations in equation \eqref{eq:doubly_deformed_def}, which are written in terms of the Hilbert stress tensor of the theory. It is easy to see from the definition of the Hilbert stress tensor,
\begin{equation}
    T_{\mu \nu} = - \frac{2}{\sqrt{-g}} \frac{\delta \mathcal{S}}{\delta g^{\mu \nu}} \, ,
\end{equation}
that the only term in (\ref{eq:semi_plus_wz}) which will contribute to $T_{\mu \nu}$ is the first term which involves $g^{\mu \nu} j_\mu^{(2)} j_\nu^{(2)}$. The two remaining terms are written in terms of $\varepsilon^{\mu \nu}$ and $\varepsilon^{ijk}$, respectively, and therefore do not couple to the metric either via an explicit metric contraction nor through implicit dependence on the measure. They are topological and they are not deformed along our combined $\TT$ and root-$\TT$ flow.

The deformed action can be written as
\begin{equation}
\label{eq:semi_plus_wz_deformed}
\begin{aligned}
    \mathcal{S}_{\text{SMWZ}}^{(\lambda,\gamma)} &= \hay\int\limits_{\partial B}\de^2\sigma\frac{1}{2} \left( L_{\text{SMWZ}}^{(\lambda,\gamma)} ( x_1, x_2 ) + \ell \, \str \left( \varepsilon^{\mu \nu} j_\mu^{(1)} j_\nu^{(3)} \right) \right)  \\
    &\qquad + \kay \int\limits_{B}\de^3\sigma \frac{1}{3} \varepsilon^{ijk} \str \left( j^{(2)}_i [ j^{(2)}_j , j^{(2)}_k ] + 3j^{(1)}_i [ j^{(3)}_j , j^{(2)}_k ]  \right) \, .
\end{aligned}
\end{equation}
where
\begin{align}
    x_1 = g^{\mu\nu}\str \left( j_\mu^{(2)} j_\nu^{(2)} \right)  , \qquad x_2 = g^{\mu\rho}g^{\nu\sigma}\str \left( j_\mu^{(2)} j_\nu^{(2)} \right) \str \left( j_\rho^{(2)} j_\sigma^{(2)} \right),
\end{align}
as before, and $L_{\text{SMWZ}}^{(\lambda,\gamma)} ( x_1, x_2 )$ is the same Lagrangian \eqref{eq:doubleformedL} that we considered above. The only feature of this Lagrangian which we will need for the present analysis is that the corresponding current
\begin{equation}
\label{eq:commutator_for_semi_wz}
    \CC{J}^{(2)}_\mu = 2\frac{\partial L_{\text{SMWZ}}^{(\lambda,\gamma)}}{\partial x_1}j^{(2)}_\mu+4g^{\nu\rho}\frac{\partial L_{\text{SMWZ}}^{(\lambda,\gamma)}}{\partial x_2}\tr[j^{(2)}_\mu j^{(2)}_\nu]\,j^{(2)}_\rho \, , 
\end{equation}
built from $L_{\text{SMWZ}}^{(\lambda,\gamma)} (x_1,x_2)$ (where $x_0$ does not play a role) satisfies 
\begin{equation}
    \big[\CC{J}^{(2)}_+,j^{(2)}_-\big]-\big[j^{(2)}_+,\CC{J}^{(2)}_-\big] =0 \, , \qquad \big[\CC{J}^{(2)}_+,\CC{J}^{(2)}_-\big]=\big[j^{(2)}_+,j^{(2)}_-\big] \, , 
\end{equation}
as we saw before, \textit{cf.}~\eqref{eq:commutatorid1sssm}--\eqref{eq:commutatorid2sssm}. In terms of this current, the equations of motion for the deformed model are
\begin{equation}
\label{eq:eom_semi_plus_wz_deformed}
\begin{aligned}
    0 &= \left[ \left( \ell \varepsilon^{\mu\nu}j^{(2)}_\mu-g^{\mu\nu}\CC{J}^{(2)}_\mu\right),j^{(3)}_\nu\right] - \frac{\kay}{\hay} \varepsilon^{\mu \nu} [ j^{(2)}_\mu , j^{(1)}_\nu ]  ,  \\
    g^{\mu\nu}D_{\mu} \CC{J}^{(2)}_\nu &=  \frac{\ell}{2} \varepsilon^{\mu\nu} \left(  \left[j^{(1)}_\mu,j^{(1)}_\nu\right] -  \left[j^{(3)}_\mu,j^{(3)}_\nu\right] \right)+ \frac{\kay}{\hay}\varepsilon^{\mu\nu} \left( [ j^{(1)}_\mu , j^{(3)}_\nu ] + [ j^{(2)}_\mu , j^{(2)}_\nu ] \right) ,  \\
    0 &= \left[ \left( \ell \varepsilon^{\mu\nu}j^{(2)}_\mu+g^{\mu\nu}\CC{J}^{(2)}_\mu \right) ,j^{(1)}_\nu \right] + \frac{\kay}{\hay} \varepsilon^{\mu \nu} [ j_\mu^{(2)} , j_\nu^{(3)} ] \, . 
\end{aligned}
\end{equation}
We now claim that these equations of motion, along with the Maurer-Cartan identities, are equivalent to the flatness of the Lax connection
\begin{equation}
\begin{aligned}
    \CC{L}_{\pm}^{(\gamma)} &= j_{\pm}^{(0)} + \ell \frac{z^2 + 1}{z^2 - 1} j_{\pm}^{(2)} \pm \left( \frac{\kay}{\hay} - \frac{2 \ell z}{z^2 - 1 } \right) \CC{J}_{\pm}^{(2)} + \left( z + \frac{\ell}{1 - \frac{\kay}{\hay} } \right) \sqrt{ \frac{\ell \left( 1 - \frac{\kay}{\hay} \right) }{z^2 - 1 } } j_{\pm}^{(1)} \nonumber \\
    &\quad + \left( z - \frac{\ell}{1 + \frac{\kay}{\hay} } \right) \sqrt{ \frac{ \ell \left(  1 + \frac{\kay}{\hay} \right) }{ z^2 - 1 } } j_{\pm}^{(3)} \, ,
\end{aligned}
\end{equation}
at any value of the spectral parameter $z$. As in the undeformed case, we will decompose this flatness condition into its components along the bosonic and fermionic subalgebras. In order to compute the commutator $[ \CC{L}_+ , \CC{L}_- ]$, it is convenient to note that the contribution arising from $j_{\pm}^{(2)}$ and $\CC{J}_{\pm}^{(2)}$, namely
\begin{align}
    &\left[ \ell \frac{z^2 + 1}{z^2 - 1} j_{+}^{(2)} + \left( \frac{\kay}{\hay} - \frac{2 \ell z}{z^2 - 1 } \right) \CC{J}^{(2)}_{+} \, , \, \ell \frac{z^2 + 1}{z^2 - 1} j_{-}^{(2)} - \left( \frac{\kay}{\hay} - \frac{2 \ell z}{z^2 - 1 } \right) \CC{J}^{(2)}_{-} \right] \nonumber \\
    &\qquad = \left( \ell^2 \left( \frac{z^2 + 1}{z^2 - 1} \right)^2 - \left( \frac{\kay}{\hay} - \frac{2 \ell z}{z^2 - 1} \right)^2 \right) [ j_+^{(2)} , j_-^{(2)}] \, , 
\end{align}
is identical to that of the undeformed theory, see eq.~\eqref{eq:bosonic_flatness_lax}. Here we have used the commutator property~\eqref{eq:commutator_for_semi_wz}. 
The bosonic part of the flatness condition for our candidate Lax connection is then
\begin{equation}
\label{eq:bosonic_flatness_lax_deformed}
\begin{aligned}
    \alg{g}_{\text{B}} : \; 0 &= F_{+-} + \left( \ell^2 \left( \frac{z^2 + 1}{z^2 - 1} \right)^2 - \left( \frac{\kay}{\hay} - \frac{2 \ell z}{z^2 - 1} \right)^2 \right) [ j_+^{(2)} , j_-^{(2)}]   \\
    &\ + \left( \left( z + \frac{\ell}{1 - \frac{\kay}{\hay} } \right) \left( z - \frac{\ell}{1 + \frac{\kay}{\hay} } \right) \frac{ \ell \sqrt{ 1 - \frac{\kay^2}{\hay^2} } }{z^2 - 1} \right) \left( [ j_+^{(1)} , j_-^{(3)} ] + [ j_+^{(3)} , j_-^{(1)} ]  \right)  \\
    &\ + \left( \ell \frac{z^2 + 1}{z^2 - 1} \right) \left( D_+ j_-^{(2)} - D_- j_+^{(2)} \right) - \left( \frac{\kay}{\hay} - \frac{2 \ell z}{z^2 - 1} \right) \left( D_+ \CC{J}^{(2)}_- + D_- \CC{J}^{(2)}_+ \right)  \\
    &\ + \left( z + \frac{\ell}{1 - \frac{\kay}{\hay} } \right)^2  \frac{ \ell \left( 1 - \frac{\kay}{\hay} \right) }{z^2 - 1 }  [ j_+^{(1)} , j_-^{(1)} ]  + \left( z - \frac{\ell}{1 + \frac{\kay}{\hay} } \right)^2  \frac{ \ell \left( 1 + \frac{\kay}{\hay} \right) }{ z^2 - 1 }  [ j_+^{(3)} , j_-^{(3)} ]  \, .
\end{aligned}
\end{equation}
Likewise, the fermionic component of the flatness condition is
\begin{equation}
\label{eq:fermionic_flatness_lax_deformed}
\begin{aligned}
    \alg{g}_{\text{F}}  : \; 0 &= \left( z + \frac{\ell}{1 - \frac{\kay}{\hay} } \right) \sqrt{\ell \left( 1 - \frac{\kay}{\hay} \right) } \left( D_+ j_-^{(1)} - D_- j_+^{(1)} \right)  \\
    &\ + \left( z - \frac{\ell}{1 + \frac{\kay}{\hay} } \right) \sqrt{ \ell \left( 1 + \frac{\kay}{\hay} \right) } \left[ \left( \ell \frac{z^2 + 1}{z^2 - 1} \right) j_+^{(2)} + \left( \frac{\kay}{\hay} - \frac{2 \ell z}{z^2 - 1} \right) \CC{J}^{(2)}_+ , j_-^{(3)} \right]  \\
    &\ + \left( z - \frac{\ell}{1 + \frac{\kay}{\hay} } \right) \sqrt{ \ell \left( 1 + \frac{\kay}{\hay} \right) }  \left[ j_+^{(3)} , \left( \ell \frac{z^2 + 1}{z^2 - 1} \right) j_-^{(2)} - \left( \frac{\kay}{\hay} - \frac{2 \ell z}{z^2 - 1} \right) \CC{J}^{(2)}_- \right]  \\
    &\ + \left( z - \frac{\ell}{1 + \frac{\kay}{\hay} } \right) \sqrt{ \ell \left( 1 + \frac{\kay}{\hay} \right) } \left( D_+ j_-^{(3)} - D_- j_+^{(3)} \right)  \\
    &\ + \left( z + \frac{\ell}{1 - \frac{\kay}{\hay} } \right) \sqrt{ \ell \left( 1 - \frac{\kay}{\hay} \right)  } \left[ \left( \ell \frac{z^2 + 1}{z^2 - 1} \right) j_+^{(2)} + \left( \frac{\kay}{\hay} - \frac{2 \ell z}{z^2 - 1} \right) \CC{J}^{(2)}_+ , j_-^{(1)} \right]  \\
    &\ + \left( \ell \frac{z^2 + 1}{z^2 - 1} \right) \left( z + \frac{\ell}{1 - \frac{\kay}{\hay} } \right) \sqrt{ \ell \left( 1 - \frac{\kay}{\hay} \right) }\\
    &\qquad\qquad\qquad\qquad\qquad\qquad\times\left[ j_+^{(1)} , \left( \ell \frac{z^2 + 1}{z^2 - 1} \right) j_-^{(2)} - \left( \frac{\kay}{\hay} - \frac{2 \ell z}{z^2 - 1} \right) \CC{J}^{(2)}_- \right] .
\end{aligned}
\end{equation}
By an explicit calculation,  substituting in the deformed equations of motion \eqref{eq:eom_semi_plus_wz_deformed} and the Maurer-Cartan identity \eqref{eq:sSSSMMaurerCartan}, one finds that \eqref{eq:bosonic_flatness_lax_deformed} and \eqref{eq:fermionic_flatness_lax_deformed} both vanish identically, assuming that we again impose the condition for $\ell$~\eqref{eq:ell_constraint} precisely as in the undeformed theory.
Conversely, one can obtain four independent equations from extracting the residues of \eqref{eq:bosonic_flatness_lax_deformed} and \eqref{eq:fermionic_flatness_lax_deformed} at $z = \pm 1$, two additional constraints from evaluating both equations at $z=0$, and one more relation from evaluating the $\alg{g}_{\text{F}}$ equation at another value of $z$, for instance $z = \ell/(1 + \frac{\kay}{\hay})$. This gives a system of seven equations relating the various $j_\mu^{(n)}$, $\CC{J}_\mu^{(2)}$, and their derivatives. The solution to this system satisfies the three equations of motion \eqref{eq:eom_semi_plus_wz_deformed} and the four Maurer-Cartan identities.
Therefore, we see that the implication goes in both directions, so the flatness of the Lax connection at any $z$ is equivalent to the equations of motion and Maurer-Cartan identities, as claimed.

\section{Conclusions and Outlook}
\label{sec:conclusions}

In this paper we have shown that a large class of classical integrable models still admit a Lax connection  after an arbitrary combination of Root-$\TT$ and (irrelevant) $\TT$ deformations.  
While this is not an indication of integrability for the deformation of \textit{all} classically integrable models, it is still a striking and non-trivial statement about the interplay of Root-$\TT$ deformations with symmetries. Moreover, having  explicit Lax connections at our disposal for a large class of models provides us with a powerful tool to study properties of classical solutions in the deformed theory, such as the value of the conserved charges under deformations. 

There are many natural and pressing questions. 
Firstly, properly establishing integrability at the classical level requires not only the existence of a Lax connection, but also its compatibility with the Poisson structure of the model. This is what allows us to construct charges that are not just conserved under time-evolution, but also mutually Poisson-commuting~\cite{Sklyanin:1980ij,Maillet:1985ec,Maillet:1985ek}. This is a natural step to truly establish classical integrability of these models (this is sometimes called ``strong integrability''). 
Naturally, it would also be interesting to study the classical integrability of even more general classes of theories such as Toda field theories, see \textit{e.g.}~\cite{Corrigan:1994nd}, or more general sigma models such as the ones that arise as Yang-Baxter deformations of (semi-)symmetric sigma models~\cite{Klimcik:2008eq,Delduc:2013qra,Kawaguchi:2014qwa,vanTongeren:2015soa} (see also~\cite{Seibold:2020ouf}). This would be a strong hint of integrability in general, and provide an even larger playground for the study of these deformations.

For the irrelevant $\TT$ deformation, it is possible to generate solutions by recasting the flow as a dynamical change of coordinates~\cite{Conti:2018tca}. It would be interesting to see if a similar construction is possible here, as this might provide a way to prove that Root-$\TT$ preserves integrability in general.
It would also be interesting to extend this construction to non-relativistic models and one-dimensional (quantum-mechanical) systems, like it was done for $\TT$, see~\cite{Cardy:2018jho,Ceschin:2020jto,Esper:2021hfq} and~\cite{Marchetto:2019yyt,Pozsgay:2019ekd,Gross:2019ach,Gross:2019uxi,Ebert:2022ehb}, respectively.
In fact, it was recently shown that a Root-$\TT$ deformation of a simple quantum-mechanical system preserves integrability too~\cite{Garcia:2022wad}. Again, it would be important to understand whether this is a general feature of such deformations.

An important outstanding question is how to quantise Root-$\TT$ theories (as well as ModMax theories~\cite{Bandos:2020jsw,Bandos:2020hgy,Bandos:2021rqy,Lechner:2022qhb,Sorokin:2021tge}). It might be worth undertaking this study on some particular examples which are integrable. For instance, we have seen in Section~\ref{sec:pcmwz:wzw} that the Root-$\TT$ deformation of a $G$-valued Wess-Zumino-Witten model is still classically integrable and conformal, but it does not enjoy the conservation of two (anti-)chiral $\alg{g}$-currents. This suggests that, at the very least, the Ka\v{c}-Moody symmetry of the quantum model is modified --- if the model remains conformal at the quantum level, which is not obvious. It would be interesting to perform an analysis similar to what was done for current-current deformations~\cite{Chaudhuri:1988qb, Forste:2003km, Borsato:2018spz} to determine the effect of the deformation.%
\footnote{%
Another direction worth exploring with the aim of quantisation, following recent work on ModMax~\cite{Lechner:2022qhb}, would be to understand the Root-$\TT$ model by adding auxiliary fields, which should make the deformed action analytic.
}

An immediate by-product of our construction is the observation that Root-$\TT$ deformations ``play well'' with symmetries. In particular, they manifestly preserve the $G$-symmetries of the sigma models that we have considered. In the case where $G$ is a supergroup, this means that they preserve integrability and \textit{target-space supersymmetry}. One of the crucial applications of the semi-symmetric sigma models has been the study of string theory on curved backgrounds --- the most celebrated example being $AdS_5\times S^5$~\cite{Metsaev:1998it,Bena:2003wd,Arutyunov:2009ga}, as well as other $AdS_n\times M_{10-n}$ backgrounds~\cite{Klose:2010ki,Sfondrini:2014via},%
\footnote{%
In fact, the introduction of the WZ term in the semi-symmetric sigma model by~\cite{Cagnazzo:2012se} has been a key ingredient in the development of integrability for $AdS_3/CFT_2$~\cite{Hoare:2013pma,Lloyd:2014bsa}.%
} 
and deformations thereof.
Even if the deformed actions are not of the sigma-model type, it would be very interesting to consider them as potential deformations of a string action. In particular, they may still enjoy a local fermionic symmetry inherited from the  $\kappa$-symmetry of the original sigma-model, which is usually related to the condition of integrability (see \textit{e.g.}~\cite{Arutyunov:2009ga,Cagnazzo:2012se}).%
\footnote{%
It is worth remarking that, even for sigma-model actions, $\kappa$-symmetry does not imply Weyl invariance of the action~\cite{Wulff:2016tju}, see \textit{e.g.}~\cite{Arutyunov:2015qva,Arutyunov:2015mqj,Borsato:2018idb,Hoare:2018ngg} for some examples. Hence, it would actually be interesting, though possibly harder, to check the stronger property of Weyl invariance.
}  
On top of the obvious physical interest of this study, this would probably allow for the perturbative quantisation of the deformed models once a suitable gauge is fixed. In fact, we would expect the gauge-fixed theory to become analytic in the small-field expansion of the transverse modes of the string (like it is the case for the Nambu-Goto action). In the case of $\TT$ deformation, a similar analysis was performed in the uniform light-cone gauge~\cite{Arutyunov:2005hd,Arutyunov:2009ga}, see e.g.~\cite{Baggio:2018gct,Frolov:2019nrr,Frolov:2019xzi,Sfondrini:2019smd}.
For Root-$\TT$ this might offer a way, however roundabout, to interpret the deformation at the level of the S~matrix (like it was done for the irrelevant $\TT$ deformation~\cite{Smirnov:2016lqw,Cavaglia:2016oda,Dubovsky:2017cnj}) and obtain some information on the deformed spectrum via the thermodynamic Bethe ansatz~\cite{Zamolodchikov:1989cf,Dorey:1996re} for a suitable mirror model~\cite{Arutyunov:2007tc}.
In a first approximation one could ask this question for bosonic strings in light-cone gauge, but it would be quite intriguing to understand it for $AdS$-type backgrounds, as it might have important implications for their holographic duals.

Another natural question is whether these deformations ``play well'' also with $(1+1)$-dimensional \textit{worldsheet supersymmetry}.
Encouragingly, it is possible to supersymmetrise the ModMax model~\cite{Bandos:2021rqy}.  Since the dimensional reduction of ModMax yields a Root-$\TT$ flow~\cite{Babaei-Aghbolagh:2022uij, Ferko:2022iru, Conti:2022egv} with a particularly simple seed action, one may suspect that supersymmetry persists for any supersymmetric seed. 
This is certainly the case for the irrelevant $\TT$ deformation~\cite{Baggio:2018rpv,Chang:2018dge,Jiang:2019hux,Chang:2019kiu, Ferko:2019oyv,Ferko:2021loo,Ebert:2022xfh}. In that case, it was useful to understand the whole deformation in terms of superfields; it may be important to do the same here. 

All the above questions are very interesting in and of themselves, but they also open the road to applying these deformations to stringy and holographic setups (both on the worldsheet and in the dual CFT) and to a deeper understanding of the ModMax theories~\cite{Bandos:2020jsw,Bandos:2020hgy,Bandos:2021rqy,Lechner:2022qhb,Sorokin:2021tge}, to which Root-$\TT$ deformations are intimately connected.

\section*{Acknowledgements}

R.B.\ and A.S.\ would like to dedicate this work to Kurt Lechner, from whom they first learned field theory.

We are grateful to Ben Hoare, Fiona Seibold, Dima Sorokin, Gabriele Tartaglino-Mazzucchelli, and Roberto Tateo for discussions related to this work.

We are grateful to the Kavli Institute for Theoretical Physics in Santa Barbara for hosting two of us (RB \& AS) during the \textit{Integrable22} workshop. The stimulating atmosphere of the program led to the discussions that eventually resulted in this work. 

The work of RB was supported by the fellowship of ``la Caixa Foundation'' (ID 100010434) with code LCF/BQ/PI19/11690019, by AEI-Spain (under project PID2020-114157GB-I00 and Unidad de Excelencia Mar\'\i a de Maetzu MDM-2016-0692),  Xunta de Galicia (Centro singular de investigaci\'on de Galicia accreditation 2019-2022, and project ED431C-2021/14), and by the European Union FEDER. C.\,F. is supported by U.S. Department of Energy grant DE-SC0009999 and by funds from the University of California. This research was supported in part by the National Science Foundation under Grant No.\ NSF PHY-1748958.
A.\,S. acknowledges support from the European Union -- NextGenerationEU, and from the program STARS@UNIPD, under project ``Exact-Holography'', \textit{A new exact approach to holography: harnessing the power of string
theory, conformal field theory, and integrable models}.

\appendix

\section{Derivation of the Equations of Motion}
\label{app:eom}
We collect here the derivation of the equations for the (deformed) models of our interest.

\subsection{Deformed Principal Chiral Model}
\label{app:eomPCM}
If we consider an infinitesimal variation of the group element $g\in G$ as $\delta g=g\epsilon$ with $\epsilon\in \mathfrak{g}$, then the variation of the Maurer-Cartan current reads $\delta j_\mu=\partial_\mu\epsilon +[j_\mu,\epsilon]$. Taking $x_1=\tr(j_\mu j^\mu)$ and $x_2=\tr(j_\mu j_\nu)\tr(j^\mu j^\nu)$ (as done in~\eqref{eq:x0x1x2PCM}) one finds
\begin{align}
    \delta x_1&=2\tr(j^\mu\delta j_{\mu})=2\tr(j^\mu(\partial_{\mu}\epsilon+[j_\mu,\epsilon]))=2\tr(j^\mu \partial_{\mu}\epsilon),
\end{align}
where we used that the term with $ [ j_\mu , \epsilon ]$ vanishes thanks to the cyclicity of the trace. Similarly, we have
\begin{align}
     \delta x_2&=4\tr(j^{\mu} j^{\nu})\tr(j_\mu \partial_\nu\epsilon).
\end{align}
It follows that the variation of the generic Lagrangian $L(x_1,x_2)$ is
\begin{align}
    \delta L &= 2 \frac{\partial L}{\partial x_1}  \tr \left(  \partial_\mu \epsilon \,  j^\mu \right) + 4 \frac{\partial L}{\partial x_2} \tr \left( j_\mu j_\nu \right) \tr \left(  \partial^\mu \epsilon \, j^\nu \right) \, .
\end{align}
When considering the variation of the action $\mathcal{S}=\int \de^2 \sigma\, L$, we can integrate by parts and conclude that 
\begin{equation}
    \delta \mathcal{S}=-\int \de^2 \sigma\,\tr\left[ \epsilon\, \partial_\mu \CC{J}^\mu\right]\,,
\end{equation}
where $\CC{J}_\mu$ is
\begin{equation}
\label{eq:currentPCMapp}
\CC{J}_\mu =2\frac{\partial L}{\partial x_1} j_\mu +
    4\frac{\partial L}{\partial x_2} \tr[j_\mu j_\nu]\,j^\nu\,,
\end{equation}
as we presented in eq.~\eqref{eq:deformedPCMeomgeneric} of the main text.

\paragraph{Adding a Wess-Zumino Term.}\label{sec:adding-wz}
When adding a WZ term to the action, of the form
\begin{align}
    \mathcal{S}_{\text{WZ}}=\frac{\kay}{6}\int \de^3\sigma\, \epsilon^{ijk}\tr\left(j_i[j_j,j_k]\right),
\end{align}
the equations of motion are modified by the additional contribution
\begin{align}
    \delta \mathcal{S}_{\text{WZ}}=\frac{\kay}{2}\int \de^2 \sigma\, \tr\left(\epsilon(\partial_+j_--\partial_-j_+)\right).
\end{align}
Therefore we have that they take the form of the conservation of
\begin{equation}
\partial_{+}\left(\hay\CC{J}_- +\kay j_-\right)+\partial_{-}\left(\hay\CC{J}_+ -\kay j_+\right)=0\,,
\end{equation}
with $\CC{J}_\mu$ as in~\eqref{eq:currentPCMapp}, that is eq.~\eqref{eq:PCMWZeomlambda} of the main text.

\subsection{Deformed Symmetric Space Sigma Model}
\label{app:eomSSSM}
In the case of the SSSM it is conventient to rewrite
\begin{equation}
   \delta j_\mu=D_\mu\epsilon +\left[j^{(2)}_\mu,\,\epsilon\right]\,, 
\end{equation}
where $D_\mu$ is the covariant derivative defined in~\eqref{eq:covariantD}. Then, writing \eqref{eq:x1x2SSSM} as 
\begin{equation}
\label{eq:x1x2appSSSM}
    x_1=g^{\mu\nu}\tr[j^{(2)}_\mu j^{(2)}_{\nu}]\,,\qquad
    x_2=g^{\mu\rho}g^{\nu\sigma}\tr[j^{(2)}_\mu j^{(2)}_\nu]\tr[j^{(2)}_{\rho} j^{(2)}_{\sigma}]\,,
\end{equation}
one obtains
\begin{align}
    \delta x_1&=2g^{\mu\nu}\tr[j^{(2)}_\mu\delta j^{(2)}_\nu]=2g^{\mu\nu}\tr[j^{(2)}_\mu\delta j_{\nu}]=2g^{\mu\nu}\tr[j^{(2)}_\mu D_{\nu}\epsilon],
\end{align}
where we used that $\tr(x^{(0)}y^{(2)})=0$ for any $x^{(0)} \in \mathfrak{g}_0$, $y^{(2)} \in \mathfrak{g}_2$, and the cyclicity of the trace, and all indices are contracted with the metric. Similarly, we have
\begin{align}
     \delta x_2&=4g^{\mu\rho}g^{\nu\sigma}\tr[j^{(2)}_{\mu} j^{(2)}_\nu]\tr[j^{(2)}_\rho D_\sigma\epsilon].
\end{align}
This is enough to conclude that the variation of a Lagrangian $L(x_1,x_2)$ is
\begin{align}
    \delta L &= 2g^{\mu\nu} \frac{\partial L}{\partial x_1}  \tr \left[  D_\mu \epsilon  \, j^{(2)}_\nu \right] + 4g^{\mu\rho}g^{\nu\sigma} \frac{\partial L}{\partial x_2} \tr \left[ j^{(2)}_\mu j^{(2)}_\nu \right] \tr \left[  D_{\rho} \epsilon \,  j^{(2)}_\sigma \right] \, ,
\end{align}
and that the variation of the action $\mathcal{S}=\int \de^2 \sigma\, L$ is
\begin{equation}
    \delta \mathcal{S}=-\int \de^2 \sigma\,g^{\mu\nu}\tr\left[ \epsilon\, \partial_\mu \CC{J}^{(2)}_\nu\right],
\end{equation}
where
\begin{equation}
    \CC{J}^{(2)}_\mu = 2\frac{\partial L}{\partial x_1}j^{(2)}_\mu+4g^{\nu\rho}\frac{\partial L}{\partial x_2}\tr[j^{(2)}_\mu,j^{(2)}_\nu]\,j^{(2)}_\rho\,,
\end{equation}
as in eq.~\eqref{eq:conservationSSMdeform} of the main text.

\subsection{Deformed Semi-Symmetric Space Sigma Model}
\label{app:eomsSSSM}
The calculations in the case of the sSSSM follow closely what was done above. In this case we want to rewrite
\begin{equation}
    \delta j_\mu = D_\mu\epsilon +\left[j^{(1)}_\mu+j^{(2)}_\mu+j^{(3)}_\mu,\,\epsilon\right]\,.
\end{equation}
Now we take $x_1$ and $x_2$ as in~\eqref{eq:x1x2appSSSM} above, and compute the variations
\begin{equation}
\begin{aligned}
    \delta x_1&=2g^{\mu\nu} \str\left[j^{(2)}_\mu D_{\nu}\epsilon+\epsilon[j^{(2)}_\mu,j^{(1)}_\nu+j^{(3)}_\nu]\right],\\
     \delta x_2&=4g^{\mu\rho}g^{\nu\sigma} \str\left[j^{(2)}_\mu j^{(2)}_\nu\right]\str\left[j^{(2)}_\rho D_{\sigma}\epsilon+\epsilon[j^{(2)}_\rho,j^{(1)}_\sigma+j^{(3)}_\sigma]\right].
\end{aligned}
\end{equation}
The Lagrangian of the sSSSM can be written as the sum $L_{\sSSSM}=L_{\sSSSM}^g+L_{\sSSSM}^\varepsilon$ where
\begin{align}
    L_{\sSSSM}^g= \frac12\, \str \left[g^{\mu\nu}j^{(2)}_\mu j^{(2)}_\nu\right],\qquad\qquad
    L_{\sSSSM}^\varepsilon = \frac12\, \str \left[\varepsilon^{\mu\nu}j^{(1)}_\mu j^{(3)}_\nu\right],
\end{align}
and after promoting $L_{\sSSSM}^g$ to $L^{g,\gamma}_{\sSSSM}(x_1,x_2)$ one finds that 
\begin{align}
        \delta S^{g,\gamma}_{\sSSSM}=-\int \de^2 \sigma\,g^{\mu\nu}\str\left[ \epsilon\, (\partial_\mu \CC{J}^{(2)}_\nu+[j^{(1)}_\mu+j^{(3)}_\mu,\CC{J}^{(2)}_\nu])\right],
\end{align} where $\CC{J}^{(2)}_\mu$ is still given by the same expression.
Finally, one also has
\begin{equation}
         \delta S^{\varepsilon}_{\sSSSM}=\frac12\int \de^2 \sigma\,\varepsilon^{\mu\nu}\str\left[ \epsilon\left(D_\mu(j^{(1)}_\nu-j^{(3)}_\nu)+\left[j^{(1)}_\mu+j^{(2)}_\mu+j^{(3)}_\mu,j^{(1)}_\nu-j^{(3)}_\nu\right]\right)\right].
\end{equation}
We conclude by noting that the equations of motion for the deformed sSSSM can be written in the form 
\begin{align}
    0=D_\mu \Lambda^\mu +\left[j_\mu,\Lambda^\mu\right],\qquad
    \text{where}\quad \Lambda^\mu= g^{\mu\nu}\CC{J}^{(2)}_\nu-\frac{1}{2} \varepsilon^{\mu\nu}\left(j^{(1)}_\nu-j^{(3)}_\nu\right).
\end{align}
Projecting this expression over the four subspaces of~$\alg{g}$ gives 
\begin{equation}
\begin{aligned}
    \alg{g}_1:&\qquad
    \varepsilon^{\mu\nu}D_{\mu}j^{(1)}_\nu = \left[\left(\varepsilon^{\mu\nu}j^{(2)}_\mu-2g^{\mu\nu}\CC{J}^{(2)}_\mu\right),j^{(3)}_\nu\right],\\[0.2cm]
    \alg{g}_2:&\qquad
    g^{\mu\nu}D_{\mu}\CC{J}^{(2)}_\nu =\frac{1}{2} \varepsilon^{\mu\nu}\left(\left[j^{(1)}_\mu,j^{(1)}_\nu\right]-\left[j^{(3)}_\mu,j^{(3)}_\nu\right]\right),\\[0.2cm]
    \alg{g}_3:&\qquad
    \varepsilon^{\mu\nu}D_{\mu}j^{(3)}_\nu = \left[\left(\varepsilon^{\mu\nu}j^{(2)}_\mu+2g^{\mu\nu}\CC{J}^{(2)}_\mu\right),j^{(1)}_\nu\right],
\end{aligned}
\end{equation}
while the equation is automatically satisfied along~$\alg{g}_0$.
Using the Maurer-Cartan equation~\eqref{eq:sSSSMMaurerCartan}, this reduces to~\eqref{eq:eomsSSSMreduceddef} and, in the undeformed case, to~\eqref{eq:eomsSSSMreduced}.

\paragraph{Adding a Wess-Zumino term.}
If we consider the action~\eqref{eq:semi_plus_wz_deformed}, we have an additonal contribution to the equations of motion, which is proportional to~$\kay$.  In particular, let us define
\begin{equation}
   \mathcal S_{\text{WZ}}= \kay \int\limits_{B}\de^3\sigma \frac{1}{3} \varepsilon^{ijk} \str \left( j^{(2)}_i [ j^{(2)}_j , j^{(2)}_k ] + 3j^{(1)}_i [ j^{(3)}_j , j^{(2)}_k ]  \right).
\end{equation}
Now, to compute the variation of $\mathcal S_{\text{WZ}}$, we cannot use the compatibility of the supertrace with the $\mathbb Z_4$ grading. One finds that the variation of the integrand can be separated into two contributions, namely a total derivative and an expression to be integrated in 3 dimensions
\begin{equation}
    \delta\mathcal S_{\text{WZ}}=\kay \int\limits_{B}\de^3\sigma\  \varepsilon^{ijk} \str \left(\partial_iB_{jk}+Z_{ijk}\right),
\end{equation}
where
\begin{equation}
    B_{ij}=\epsilon^{(2)}([j^{(2)}_i ,j^{(2)}_j ]+[j^{(1)}_i ,j^{(3)}_j ])+\epsilon^{(1)}[j^{(3)}_i ,j^{(2)}_j ]+\epsilon^{(3)}[j^{(1)}_i ,j^{(2)}_j ].
\end{equation}
Importantly, $Z_{ijk}$ identically vanishes thanks to the Maurer-Cartan identity and the Jacobi identity on the superalgebra. This is crucial in order to have a well-defined WZ term, yielding equations of motion that only depend on the boundary (i.e.~2-dimensional) degrees of freedom. To conclude, the equations of motion receive the additional contribution of $\kay \varepsilon^{\mu\nu}B_{\mu\nu}$, and they read as in~\eqref{eq:eom_semi_plus_wz_deformed}.

\bibliographystyle{utphys}
\bibliography{master}

\end{document}